\documentclass[manuscript, nonacm]{acmart}
\usepackage{graphicx}
\usepackage{soul}
\usepackage{lineno}
\usepackage{tabularx}
\usepackage{multirow}
\usepackage{multicol}
\usepackage{textcomp}
\usepackage{colortbl}
\usepackage{numprint}
\usepackage{textcomp}
\usepackage{booktabs}
\usepackage{enumitem}
\usepackage{textcomp}
\usepackage[utf8]{inputenc}
\usepackage{pdfpages}
\usepackage{array}
\usepackage{float}
\usepackage{subfig}
\usepackage{multirow}
\usepackage{color}
\usepackage{algorithmic}
\usepackage{graphicx}
\usepackage{textcomp}
\usepackage{xcolor}
\usepackage{xspace}
\usepackage{geometry}
\usepackage[normalem]{ulem}
\usepackage[framemethod=default]{mdframed}
\usepackage[normalem]{ulem}
\useunder{\uline}{\ul}{}
\usepackage{adjustbox}
\usepackage{pdflscape}
\usepackage{afterpage}

\newcolumntype{P}[1]{>{\centering\arraybackslash}m{#1}}

\newcommand{\CHANGED}[1]{#1}

\newcommand{\CHANGEDFINALVERSION}[1]{#1}

\AtBeginDocument{%
  \providecommand\BibTeX{{%
    \normalfont B\kern-0.5em{\scshape i\kern-0.25em b}\kern-0.8em\TeX}}}

\newcommand{\tabitem}{~~\llap{\textbullet}~~}

\renewcommand\footnotetextcopyrightpermission[1]{} % removes footnote with conference info
\setcopyright{none}
\begin{document}

\title{The ``Shut the f**k up'' Phenomenon: Characterizing Incivility in Open Source Code Review Discussions}


\author{Isabella Ferreira}
\email{isabella.ferreira@polymtl.ca}
\affiliation{%
  \institution{Polytechnique Montréal}
  \city{Montréal}
  \country{Canada}
}

\author{Jinghui Cheng}
\email{jinghui.cheng@polymtl.ca}
\affiliation{%
  \institution{Polytechnique Montréal}
  \city{Montréal}
  \country{Canada}
}

\author{Bram Adams}
\email{bram.adams@queensu.ca}
\affiliation{%
  \institution{Queen's University}
  \city{Kingston}
  \country{Canada}
}

% \renewcommand{\shortauthors}{Ferreira et al.}

% OLD LINK FOR BLIND REVIEW
% \newcommand{\replicationPackage}{\url{https://figshare.com/s/e3520a03e91ded919834}}

\newcommand{\replicationPackage}{\url{https://doi.org/10.6084/m9.figshare.14428691.v1}}

\begin{abstract}
Code review is an important quality assurance activity for software development. Code review discussions among developers and maintainers can be heated and sometimes involve personal attacks and unnecessary disrespectful comments, demonstrating, therefore, incivility. Although incivility in public discussions has received increasing attention from researchers in different domains, the knowledge about the characteristics, causes, and consequences of uncivil communication is still very limited in the context of software development, and more specifically, code review. To address this gap in the literature, we leverage the mature social construct of incivility as a lens to understand confrontational conflicts in open source code review discussions. For that, we conducted a qualitative analysis on 1,545 emails from the Linux Kernel Mailing List (LKML) that were associated with rejected changes. We found that more than half (\CHANGED{66.66\%}) of the non-technical emails included uncivil features. Particularly, \textit{frustration}, \textit{name calling}, and \CHANGED{\textit{impatience}} are the most frequent features in uncivil emails. We also found that there are civil alternatives to address arguments, while uncivil comments can potentially be made by any people when discussing any topic. Finally, we identified various causes and consequences of such uncivil communication. Our work serves as the first study about the phenomenon of in(civility) in open source software development, paving the road for a new field of research about collaboration and communication in the context of software engineering activities.
\end{abstract}




\maketitle

\section{Introduction}

Code review is a software quality assurance practice widely adopted in open source software projects~\cite{Bacchelli2013}. In this practice, a \textit{developer} submits a collection of functionally coherent changes to the source code (i.e., a \textit{patch}) that implements a new feature or fixes a bug for review; project \textit{maintainers} (a group of core community members) are then responsible for reviewing the patch and making the decision of either integrating it into the project (patch acceptance) or not (patch rejection). At the core of the code review process is the discussion among the developers and maintainers around various topics related to the submitted patches; topics such as the merits of the patch to the project, the appropriateness of the solution design, and its implementation details, to name a few. During such discussions, the maintainers frequently ask for clarifications, offer suggestions, and provide critical feedback to the developers, while the developers explain their rationale, hoping to convince the maintainers to accept the patch. The review process and the embedded discussions are usually supported by tools such as Gerrit~\cite{hamasaki2013does} and mailing lists~\cite{tourani2014monitoring, jiang2014tracing}.

While both the maintainers and the contributing developers may have the same intention of enriching functionality and fixing problems for the software project, they can have different or even conflicting interests around the code review activity. Particularly, for open source software projects, developers can have diverse motivations in providing code contribution~\cite{alexander2002working, hertel2003motivation}. In addition to intrinsic motivations such as community identification and the desire of feeling competent, researchers have identified various extrinsic motivations of contributing to open source projects such as self-marketing, gaining revenue for related products, and personal needs for specific software functionalities~\cite{alexander2002working}. However, project maintainers face the daily gate-keeping challenges of not only ensuring the quality of the software but also determining the relevance of the proposed functionalities to the greater community. As a result, while developers always hope to make their code contributions accepted, it is the maintainers' responsibility to critically assess each contribution, candidly communicate with the developers, and make informed decisions. Because of this nature of code review discussions, they innately contain disagreements and potential conflicts.

This situation is exacerbated by several factors related to large-scale open source software projects, such as Linux. Because of the reputation and wide adoption, the sheer amount of contributions received in those communities can be overwhelming. For example, the daily volume of emails sent to the Linux Kernel Mailing List is more than 1,000~\cite{dailyemailslkml}; a typical maintainer receives hundreds of emails per day~\cite{Tan2019}. The quality and focus of these contributions are also often very diverse, which adds to the burden of the maintainers~\cite{Tan2020}, resulting in stressful and non-effective communication. Further, the power differences between maintainers and developers are evident in popular open source projects, creating an uneven platform for discussion. In our preliminary investigation of code review of Linux kernel development, we have found that the discussions can be heated and sometimes involve personal attacks and unnecessary disrespectful comments; i.e., the code review discussions can demonstrate \textit{incivility}. We can see it very clearly from the following code review comment of a patch submitted to the Linux kernel: %, in which a maintainer comments on the review of another maintainer about a bug that breaks user space (a set of locations where everything other than the kernel run):

% https://lkml.org/lkml/2012/12/23/75

\begin{quotation}
\textit{``
[Person's name], SHUT THE F**K UP! ... % It's a bug alright - in the kernel. How long have you been a maintainer? And you *still* haven't learnt the first rule of kernel maintenance? [...] We never EVER blame the user programs. How hard can this be to understand? Shut up, [Person's name]. And I don't\_ever\_ want to hear that kind of obvious garbage and idiocy from a kernel maintainer again. Seriously... And you've shown yourself to not be competent in this issue, so I'll apply it directly and immediately myself. Seriously. How hard is this rule to understand? We particularly don't break user space with TOTAL CRAP. I'm angry, because 
your whole email was so \_horribly\_ wrong, and the patch that broke things was so obviously crap. Fix your f*cking "compliance tool", because it is obviously broken. And fix your approach to kernel programming.''}
\end{quotation}

%Civility is the minimum degree of politeness and courtesy required in a social situation~\cite{dictionary}. Whereas a decade ago civility was focused on political and religious debates, nowadays civility focuses instead on how individuals speak to each other and, more particularly, how they are able to disagree~\cite{bejan2017mere}. 

Incivility in public discussions has received increasing attention in recent years. Researchers have investigated this phenomenon in the domains of interpersonal relationships in workplace dynamics~\cite{blau2005testing, forni2010choosing}, political discourse~\cite{brooks2007beyond, fridkin2008dimensions, bejan2017mere}, and online comments~\cite{coe2014online,maity2018opinion,molina2018role}, to name a few. According to Bejan~\cite{bejan2017mere}, incivility is the product of technology, social, and cultural transformations unique to the modern world. That is, with the increasing opportunities for public debates on prevalent platforms such as social media, Q\&A systems, and tools for remote and collaborative work, incivility can spread more rapidly and widely than ever before~\cite{sobieraj2011incivility}. 

In the context of software development in general and code review discussions in particular, however, our knowledge about the characteristics, causes, and consequences of uncivil communication is still very limited. Many studies have investigated the technical aspects of code review, such as the kinds of patches that are more likely to be accepted~\cite{jiang2013will}, the reasons why patches are rejected~\cite{tao2014writing}, the characteristics of the reviewing history in mailing lists~\cite{jiang2014tracing}, and the techniques for recommending appropriate reviewers for patches~\cite{jeong2009improving}. The social aspects of code review discussions, however, are only explored in a few very recent studies~\cite{Bosu2017,egelman2020predicting}, often not directly leveraging the construct of incivility. For example, Egelman et al.~\cite{egelman2020predicting} have found that unnecessary interpersonal conflicts in code review can evoke negative feelings, such as frustration and discouragement, in developers. 

In this paper, we thus contribute to the first study that leverages the mature social construct of incivility as a lens to understand confrontational conflicts in open source code review discussions. Particularly, we focus on identifying the discussion characteristics, the causes, and the consequences of uncivil code review discussion comments from both maintainers and developers. 
% Unfortunately, not many tools nor many practices exist to mitigate incivility. In Software Engineering (SE), the most common practice is the code of conduct, which establishes expectations for communication and behavior between community members~\cite{tourani2017code}. Its goal is to make everyone comfortable in contributing to the project and to foster a healthy environment~\cite{tourani2017code}.
To achieve our goals, we conducted a qualitative analysis on 1,545 emails from the Linux Kernel Mailing List (LKML) that were associated with rejected patches; we study rejected patches because the rejection could be the first indication of conflict, inducing incivility, and thus would allow us to achieve an understanding of both uncivil and civil ways of addressing conflicts. \CHANGED{Furthermore, previous work has shown that rejected patches represent more than 66\% of all patches submitted to LKML~\cite{jiang2013will}, and that the Linux community frequently rejects patches using a harsh language when reporting the rejection, even though the reasons for rejection are purely technical~\cite{alami2019does}}. %We characterize incivility based on tone-bearing discussing features (TBDF), which are sentences that convey an unnecessarily disrespectful tone towards the discussion forum, its participants, or its topics~\cite{coe2014online}. %We aim at identifying the TBDF present in discussions of rejected patches, analyzing the percentage of review e-mails that are uncivil, analyze how incivility is correlated with arguments, contributors, and topics of the discussions, and finally analyze the causes and impacts of uncivil communication. 
Overall, by analyzing the discussions around rejected patches on LKML, we aimed at answering five research questions and generated the following contributions.

\vspace{6pt}
\noindent
\textbf{RQ1: Which features of discussion can be found in code reviews of rejected patches?}

We focused on identifying tone-bearing discussing features (TBDF) in code review comments. We define TBDF as \textit{conversational characteristics demonstrated in a written sentence that convey a mood or style of expression}. This concept was inspired from Coe et al.'s definition of incivility as ``features of discussion that convey an unnecessarily disrespectful tone''~\cite{coe2014online}. We identified \CHANGED{16} TBDFs that emerged from an inductive analysis of code review discussions of rejected patches, including \CHANGED{seven} uncivil features: \textit{bitter frustration}, \CHANGED{\textit{impatience}}, \textit{irony}, \CHANGED{\textit{mocking}}, \textit{name calling}, \CHANGED{\textit{threat}}, and \textit{vulgarity}.% We could identify TBDF that convey either positive, neutral, negative, or uncivil tone.\\

\vspace{6pt}
\noindent
\textbf{RQ2: How much incivility exists in code review discussions of rejected patches?}

We found that although the majority of the code review emails did not contain a TBDF (i.e., focused on technical discussions), more than half (\CHANGED{66.66\%}) of the non-technical emails included uncivil features. \textit{Frustration}, \textit{name calling}, and \CHANGED{\textit{impatience}} are the most frequent features in uncivil emails.

\vspace{6pt}
\noindent
\textbf{RQ3: How is incivility correlated with the occurrence of arguments, the individual contributors, and the discussion topics?}

We aimed to explore these correlations to identify potential explanations of uncivil communication. However, we did not find evidence that incivility was associated with any of the three attributes. These results indicated that there are civil alternatives to address arguments, while uncivil comments can potentially be made by any people when discussing any topic. As a result, incivility might have been triggered by other factors, which we explore in RQ4.

%Against our expectations, almost half (46.79\%) of the uncivil emails were part of threads without an argument. Furthermore, 7 individuals have sent only uncivil emails, while 50 people have sent not only uncivil emails but also civil and technical emails. We found that there is no difference in the number of civil and uncivil emails sent by someone who has sent at least one uncivil emails. Finally, we found that both developers and maintainers are often uncivil on emails discussions about \textit{workflow}, however, we could not identify a topic that most cases of incivility happened. Thus, we conclude that incivility is not related to arguments, contributors, and topics of thediscussion.\\

\vspace{6pt}
\noindent
\textbf{RQ4: What are the discoursal causes of incivility?}

Through examining discussions before the uncivil comments, we identified \CHANGED{eight} themes that caused incivility for developers and five themes for maintainers. \textit{Violation of community conventions} \CHANGED{was the common cause} of incivility for both developers and maintainers. Further, maintainers were also frequently irritated by \textit{inappropriate solution proposed by developer}, while developers by characteristics in \textit{the reviewer's feedback}.

\vspace{6pt}
\noindent
\textbf{RQ5: What are the discoursal consequences of incivility?}

Through examining discussions after the uncivil comments, we identified \CHANGED{eight} themes as consequences of uncivil comments made by developers \CHANGED{and by maintainers}. We found that most frequently, the target of the uncivil comments \textit{discontinued further discussion}; in some cases, the target continued the discussion in a civil way, while a few escalated the uncivil communication. 

% \hl{[Need a paragraph for: What are the impacts of this study? this is the first study about incivility -- pave the road for future studies about incivility in software-related collaborations and discussions -- implications to tool support for collaborative code review, community building, discussion, etc.]}

Our results characterize the civil and uncivil comments in open source code review discussions and support the notion that open source communities might be able to create healthier and more attractive environments by fostering civil arguments. Concretely, based on the uncivil TBDFs we identified, if code review discussion participants cease the expression of bitter frustration, name calling, and \CHANGED{impatience}, reviews and arguments might be more constructive and efficient. Overall, our paper makes the following contributions:

\begin{itemize}
    \item Our effort serves as a first study about in(civility) in open source communities, therefore, paving the road for future studies about this topic in software-related collaborations and discussions.
    %\item An in-depth characterization of incivility in code review discussions by identifying the TBDFs of uncivil emails as well as the causes and consequences of uncivil communication on both developers and maintainers.
    \item \CHANGED{We provided an in-depth characterization of incivility in open source code review discussions, providing evidence, descriptions, and explanations of incivility in this dynamic context. By analyzing the code review discussions in the Linux Kernel Mailing List, we encountered TBDFs not previously found in any other study, proposed a definition of incivility based on the uncivil TBDFs, assessed the frequency of incivility, analyzed the correlation with the common assumptions of the cause of incivility (i.e., arguments, contributors, and topics), and assessed the causes and consequences of developers' and maintainers' uncivil interactions.}
    
    \item We suggested practical implications and tool design ideas proposed \CHANGED{to open source software communities and researchers}, encouraging future efforts to help software communities address incivility and create healthy working environments.
\end{itemize}
\section{A Motivational Case Study: How do Open Source Contributors Perceive Incivility?}
\label{sec:case-study}

Since little is known about incivility in the context of open source software development, this section explores the perceptions of this concept from members of open source communities. Particularly, we conducted surveys and an open discussion with open source community members during a Birds of a Feather (BoF) session at the Linux Plumbers Conference\footnote{\url{https://linuxplumbersconf.org/event/4/contributions/543/}}. BoF sessions are informal gatherings of people interested in a particular topic during industrial conferences.

\subsection{Case study methods}
The goal of the BoF was to raise awareness about civility and to learn from the community how their communication happens in practice as well as the associated challenges. The BoF session focused on \CHANGED{four} main topics: understanding what civility means to the participants, discussing incivility in open source communities, discussing the role of the code of conduct, and evaluating whether it is feasible \CHANGED{for sentiment analysis tools} to automatically detect incivility. \CHANGED{An intuitive idea was to identify incivility through sentiment analysis tools. Hence, we wanted to assess if the tools' results are compatible with the Linux contributors' interpretation of (in)civility}.

The BoF session lasted 45 minutes. We first presented the concept of civility, then we brought up a survey whose questions were interspersed with group discussions. The participants had about ten minutes to answer the online survey \CHANGED{composed of 14 closed-ended questions on the participants' experience with uncivil communication, the code of conduct, and the contributors' perceptions of (in)civility in three communication examples}. After everyone has submitted their answers to the survey, we displayed the anonymous answers and discussed each topic for about five to ten minutes. We had one person in the audience taking notes about the participants' discussion. After the BoF session, we conducted a thematic analysis on our notes and the survey answers to identify prominent themes discussed by the participants. \CHANGED{The survey questions and the presentation used to guide the group discussions are hosted online\footnote{\replicationPackage} for replication or third party reuse}.

In the survey, we asked the participants to discuss if they have ever experienced incivility in open source software communities, if they tried to call out the uncivil person, if they would talk to the person offline in case of an uncivil interaction, the major factors and consequences that can make communication uncivil, and to what extent communication helps to achieve civil communication goals. \CHANGED{Additionally, we asked the participants to classify three code review emails and we compared their classification with the results of off-the-shelf sentiment analysis tools}.

\CHANGED{To identify sentiment in code review discussions, we used IBM Watson~\cite{ibmwatson}, a general-purpose sentiment analysis tool, and  Senti4SD~\cite{calefato2018sentiment}, a tool developed to identify sentiment in software engineering artefacts. We run both tools in the code review emails from the Linux Kernel Mailing List (LKML), and we randomly picked one email classified by Senti4SD as positive, negative, and neutral. Since sentiment analysis tools might have disagreements, we picked the results of Senti4SD that is trained on software engineering data. During the BoF session, participants could classify the code review discussions into \textit{civil}, \textit{uncivil}, and \textit{I don't know}, in case they were not sure about the classification.}

%During the BoF session, we advised the participants that \textit{civil} emails would be the ones that express \textit{positive} sentiment, and \textit{uncivil} emails the ones that express \textit{negative} sentiment. Additionally, participants could classify the email as \textit{neutral} if they only contain technical information or the lack of sentiment, or they could classify the email as \textit{I don't know}, in case they are not sure about the classification.}

We collected survey responses from 22 participants at the BoF session; 20 provided demographic information. Among the participants, 17 were from the Linux kernel community, two were from both the Linux and Debian community, and one person was from another open source community. Seven people had from 10 to 20 years of experience contributing to open source projects, six of them had from five to ten years, five had from zero to five years of experience, and two people had more than 20 years of experience. Ten self-reported as software developer, six as maintainer, two as both developer and maintainer, and two as open source software manager. 

\subsection{Summary of findings}
\label{sec:survey-civi}
% P1: Brendan
% P2: Dan
% P3: Shuah
% P4: Frank
% P5: Greg

When developers were asked about the extent to which civil communication helps to achieve communication goals, 81.9\% mentioned either \textit{to a great extent} or \textit{to a moderate extent}. Participants discussed several factors that they considered to be associated with the concept of civility, such as \textit{no personal attacks}, \textit{distinguishing the target from the problem}, \textit{respect}, \textit{politeness}, \textit{constructive feedback}, \textit{accepting mistakes}, and \textit{being humble}. Most participants (18 out of 22) mentioned that they have experienced incivility themselves. 

Participants discussed various strategies to respond to uncivil online communication. Nine participants have recalled situations where they tried to call out the offending person online, while 14 discussed that they would talk to the person offline in case they face uncivil interactions. Participants mentioned that whether to call out a person depends on the degree of familiarity with that person (participant P1), the perceived power imbalance (P5), and the topic and the target associated with the uncivil comment (P2). Alternatively, some participants believed that a better way to handle uncivil comments is to \textit{``take the punch and get back to reality''} (P2). P4 added that normally people keep escalating the problem if nobody recedes. 

%P3 mentioned having maintained a positive relationship with people by not taking things personally.

When discussing factors that can make communication uncivil, participants mentioned several prominent themes, including (1)~differences in perception and viewpoint between the people engaged in the conversation ($N=12$), (2)~the sentiment/emotion of a discussion participant towards the source code, the topic, or the people involved in the conversation ($N=12$), and (3)~language and culture differences that led to communication barriers ($N=12$). Participants also perceived the consequences of uncivil communication to be (1)~worsened the reputation of the community and reduced attraction to new contributors ($N=19$), (2)~reduced retention, leading to contributors leaving the community ($N=17$), (3)~\CHANGED{contributors demonstrating frustration} ($N=11$), (4)~contributors showing signs of passive-aggressive behavior by indirectly expressing negative feelings ($N=10$), and (5)~patch rejection ($N=9$).

\CHANGED{The results of the code review emails classification were surprising. The first email was classified as \textit{negative} by both tools. However, 50\% of the participants classified this email as \textit{positive/civil}, 30\% as \textit{negative/uncivil}, and 20\% of the participants were not sure about the classification. During the group discussion, some participants mentioned that the classification of this first email depends on the context, i.e., on the previous emails of the email thread. Coincidentally, the recipient of this email was in the audience, so he could explain the context of this specific email. He mentioned that this email was sent to him by a co-worker (of 20 years) who was a non-native English speaker. According to him, the context, familiarity, mother tongue, and culture all matter for determining the civility of an email. He also mentioned that there was a newcomer in the same email thread, and the way to convey the message was not clear enough for the newcomer, and he needed to repeat his message many times.

The second email was classified by Senti4SD as \textit{positive} and by IBM Watson as \textit{negative}. 80\% of the participants classified the email as \textit{positive/civil}, whereas 15\% thought it was \textit{negative/uncivil}, and 5\% was not sure. The participants mentioned that \textit{granularity} matters and people might have different opinions based on the granularity of the analysis. For example, this email started as \textit{negative} and then became \textit{positive}.

Finally, the third email was classified as \textit{neutral} by Senti4SD and as \textit{negative} by IBM Watson. However, 55\% of the participants classified such email as \textit{positive/civil}, 25\% as \textit{negative/uncivil}, and 20\% as \textit{I don't know}. In this example, participants reinforced the need to understand the context of the email to perform the classification.}

\subsection{Lessons learned from the case study}

Through the small-scale survey and the BoF discussion, we found that incivility can be an important issue affecting many open source contributors in various ways. Uncivil communication can originate from diverse sources and can have wide impacts. More specifically, in the context of code review where differences in perception and viewpoint often happen, discussions have a potential for arguments, and therefore, they might be uncivil. In the current context, the most commonly adopted strategy of addressing uncivil communication seemed to be either to ``call it out'' or to ``eat it up.'' 

\CHANGED{Furthermore, the sentiment analysis results demonstrate a lack of agreement between tools (Senti4SD and IBM Watson), and between tools and humans, which shows that existing sentiment analysis tools may not be able to identify uncivil communication. Finally, participants gave us valuable insights on factors that need to be considered to better assess civility in a text, such as the context, the familiarity among people, the mother tongue and culture, and the granularity of analysis.} These findings have inspired us to \CHANGED{conduct a manual qualitative analysis on} the civil and uncivil communication styles in code review discussions to understand the phenomenon of incivility in the software engineering context. Our research questions \CHANGED{and study design} were also framed based on the information we gathered from this case study.

\section{Background \& Related Work}

In this section, we provide background information and discuss related work in the areas of (1) the modern code review process in general, (2) the code review process of the Linux kernel development, (3) the social construct of incivility, and (4) previous studies focused on negative communication in open source software development.

\subsection{The modern code review process}

Code review is a widely-adopted software engineering practice in both open source and proprietary software projects~\cite{Bacchelli2013}. In such a practice, before new changes to the code base are incorporated into the project, someone other than the author of the changed source code manually inspects the changes, provides feedback for improvements, and decides if the new changes should be applied to the code base. Practically, the changes are often combined into functionally coherent \textit{patches} for review. The modern code review process is usually lightweight, informal, and relies on code review tools and platforms, such as Gerrit\footnote{https://www.gerritcodereview.com} and ReviewBoard\footnote{https://www.reviewboard.org}. In large scale open source projects, code contribution is frequently done by various \textit{developers} who are interested in the project and the community~\cite{Cheng2019}, while code review is usually conducted by a group of trusted developers who are the core community members, sometimes called project \textit{maintainers}~\cite{Rigby2008,Thongtanunam2017}.

The code review practice does not only ensure the quality and integrity of the software being developed, but it also considerably impacts the dynamics and relationships among members in open source communities. For example, Bosu et al. have identified that code reviews have several non-technical benefits such as knowledge sharing and relationship building; they also found that carelessness and lack of respect can create negative perceptions in both developers and maintainers and hinder collaboration~\cite{Bosu2017}. Similarly, Asri et al. identified that negative sentiments expressed in code reviews were associated with prolonged issue fixing time~\cite{el2019empirical}. Ebert et al. also found that communication barriers in the code review process, originated from factors such as unarticulated rationale and lack of context, can result in lengthy discussions and delay in decision-making~\cite{Ebert2019}. Through analyzing code review comments, Pascarella et al. have identified the dynamics of reviewers' different informational needs across the life-cycle of a code review that can be satisfied by better communication and tool support~\cite{Pascarella2018}. Henley et al. have proposed an automated collaboration tool that can improve communication and productivity in code review~\cite{henley2018cfar}. \CHANGED{Alami et al. found that open source contributors often experience rejection and negative feedback, and, as a consequence, they need to take the frequent negative feedback as an opportunity to learn and to improve in their job~\cite{alami2019does}.} Our work builds upon this body of literature and focuses on identifying characteristics of uncivil communication in code review. 

\subsection{Code review in the Linux kernel development}

\begin{figure*}[ht] 
\subfloat{\includegraphics[clip, width=\linewidth]{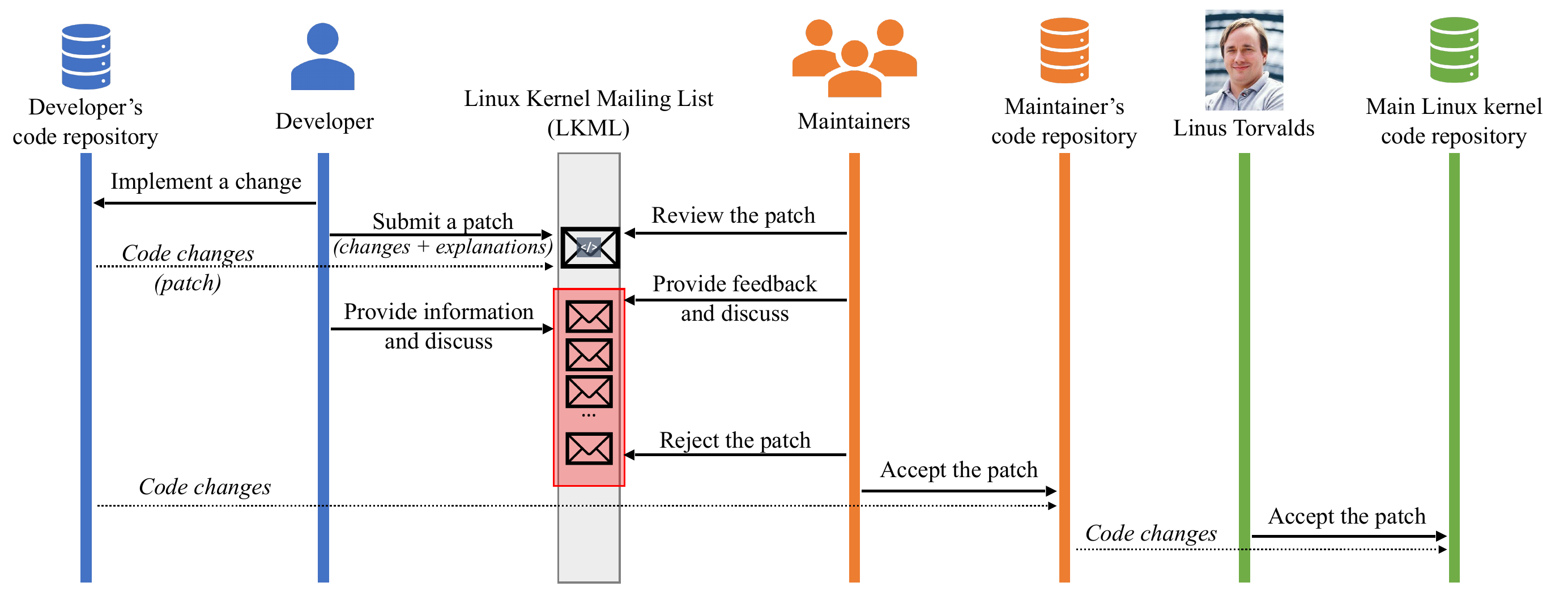}}
\caption{Summary of the code review process of the Linux kernel development.}
\label{fig:linux_code_review_process} 
\end{figure*}

Instead of using a dedicated code review tool, the Linux kernel community uses mailing lists for code review~\cite{jiang2014tracing, ferreira2019longitudinal}. The Linux kernel review process happens in the following way (see Figure~\ref{fig:linux_code_review_process} for a summary). First, a \textbf{\textit{developer}} %sends a Request for Comments (\textbf{\textit{RFC}}) to a kernel subsystem's mailing list to get feedback about ideas for new features or bug fixes. Based on the received feedback, the developer 
implements the new feature or the bug fix in their local version control system. Once finished, the developer will generate a summary of changes based on a series of commits, which is formulated as a \textbf{\textit{patch}}. The developer then submits the patch to the Linux Kernel Mailing List (LKML) or a subsystem mailing list via an \textbf{\textit{email}}; the majority of the patch submissions and the discussions and debates about the Linux kernel take place on LKML~\cite{gallivan2001striking, love2010linux}. Once a patch is submitted, the \textbf{\textit{maintainers}} will then review the patch and make one of the following three decisions: (1) accepting the patch as is, (2) rejecting the patch immediately (e.g., because the feature implemented is not interesting), or (3) provide feedback to the developer through email discussions. Our research focuses on the discussions that happened in the latter case. The discussion about the proposed patches can involve all developers and maintainers in the community. Such a discussion can be done through several email replies, which compose an \textbf{\textit{email thread}}. As a result of the discussion, the maintainers may decide to accept the patch eventually or ask the developer to send a new patch version with modifications. Once a patch is accepted, the maintainers will then \textbf{\textit{commit}} the code changes to their version control systems. Linus Torvalds, the creator of the Linux kernel, will then eventually review the patch and make the final decision of whether the patch would be included as a part of an official Linux kernel release.

\subsection{Incivility}
% Civility
% \cite{bejan2017mere, sobieraj2011incivility, , kenski2020perceptions, sadeque2019incivility, sifianou2019politeness, coe2014online}

% Civility - interpersonal relationship and work dynamics
% \cite{blau2005testing, forni2010choosing}

% Civility - political discourse
% \cite{brooks2007beyond, fridkin2008dimensions}

% Civility Twitter
% \cite{maity2018opinion}

% Civility Facebook
% \cite{molina2018role}

% Civility SE
% \cite{egelman2020predicting}

Many authors~\cite{kenski2020perceptions, coe2014online, sadeque2019incivility} believe that ``incivility is in the eye of the beholder''; in other words, what seems uncivil for one person might strike another person as completely appropriate.
Although incivility, and the converse phenomenon of civility, has been studied by many authors from different fields, there is no common agreement on its definition. For some, civility is interchangeably related to politeness. However, according to Bejan~\cite{bejan2017mere}, although civility is certainly associated with politeness, calling someone uncivil is far worse than impolite. For Bejan, impoliteness can be tolerated in a way that incivility cannot. When studying political discussion groups, Papacharissi~\cite{papacharissi2004democracy} found that relating civility to only politeness makes us ignore the democratic merit of a heated discussion. Hence, Papacharissi suggested identifying civil behaviors that enhance a democratic conversation. In the field of electorate, Brooks and Geer~\cite{brooks2007beyond} defined civility in terms of mutual respect. When studying incivility on social media, Maity et al.~\cite{maity2018opinion} defined it as \textit{``an act of sending or posting mean text messages intended to mentally hurt, embarrass or humiliate another person using computers cell phones, and other electronic devices''}. Focusing on online comments on news reports, Coe et al.~\cite{coe2014online} suggested a general definition of incivility, as \textit{``features of discussion that convey an unnecessarily disrespectful tone toward the discussion forum, its participants, or its topics.''} According to their definition, incivility is unnecessary, since it does not add anything constructive to the discussion. Our work builds on the definition proposed by Coe et al.~\cite{coe2014online} by focusing on incivility in code review discussions of rejected patches. 

The presence of incivility in public discourse has important consequences. For example, Anderson et al.~\cite{anderson2014nasty} found that people who are exposed to uncivil deliberation in blog comments of science-related blog posts are more likely to perceive the technology as risky than those who are exposed to civil comments. Consequently, the effects of user-to-user incivility on perceptions towards emerging technologies may be a problem for science experts that rely on public acceptance of their technology. In a recent work, Kenski et al.~\cite{kenski2020perceptions} found that the audience's perceptions of incivility are not uniform. For example, females usually have greater sensitivity to incivility than men, so they are less likely to engage in such discourses. Similarly, Molina et al.~\cite{molina2018role} found that users who were exposed to civil comments on Facebook were more prone to engage in discussions. Hence, instigating civil discussions can make political and social debates more likely to occur, since it sparks arguments that can be exchanged in a constructive way. Our study extends this related literature to investigate the causes and consequences of incivility in open source code reviews. \CHANGED{To the best of our knowledge, incivility has not yet been studied in the software development context}. 
%\bram{there was the last year ICSE paper? that was only about conflics, take a look at reference egelman2020predicting in the paragraph below}

\subsection{Negative communication in open source software development}
While we have not found previous studies that are directly focused on the construct of incivility in software development, researchers in software engineering and computer-supported cooperative work have investigated the effect of negative communication in the open source software development process. Researchers have identified that the negative communication styles have affected a wide range of open source developers, from newcomers who experienced communication barriers when onboarding to open source software projects~\cite{steinmacher2013newcomers} to frequent contributors who often suffer from stress and burnout facing toxic interactions~\cite{ramanstress}. \CHANGED{Furthermore, previous work has found that there are quantifiable differences in the communication patterns between leaders of the Linux kernel community, in which some people often use words such as \textit{thanks} and \textit{sorry}, while other people use rude and offensive words ~\cite{schneider2016differentiating}.} To help address the effects of negative communication, Tan and Zhou~\cite{Tan2019} identified 17 strategies as effective practices for communication when submitting a patch to LKML.

Most closely related to incivility, several previous studies focused on the concept of \textit{conflict} in open source software development. Filippova et al.~\cite{Filippova2015,Filippova2016} have conducted a series of studies to understand the types, sources, and effects of conflict in open source development from the contributors' perspectives. They have identified that conflicts are often associated with disagreements on technical tasks, development processes, and community norms; these conflicts have impacted the developers' perception of team performance and their community identification, which in turn influenced their intention to remain in the project. Focusing on the effectiveness of conflict management strategies, Huang et al.~\cite{Huang2016} found that only the strategy of providing concrete constructive suggestions for alternative suggestions at the technical level was effective in reducing the negative consequences of conflicts, while neither rational clarifications of misunderstandings nor social encouragement were effective. Most recently, Egelman et al.~\cite{egelman2020predicting} identified five types of feelings (e.g., frustration and discouragement) that were results of \textit{``unnecessary interpersonal conflict''} in code review at Google. They further developed a method that uses metrics in the code review process (i.e., rounds of review, review time, and shepherding time) to predict these feelings. Different from these studies, we leverage a mature social construct of incivility as a lens for a systematic understanding of the characteristics, causes, and consequences of the most detrimental type of conflict in open source software development.

\section{Methods}

This section discusses the case study approach used by this paper on the Linux Kernel Mailing List in order to characterize incivility in code review discussions of rejected patches.

\subsection{Collecting code review emails}
\label{sec:data-collection}

% Get all individual emails from 2018/01/01 - 2019/03/20 from LKML

We collected code review emails from the Linux Kernel Mailing List (LKML) in the period between January 2018 and March 2019. We chose to study the Linux community because they have a diverse and large number of contributors with different communication styles~\cite{schneider2016differentiating} as well as a large number of daily discussions~\cite{dailyemailslkml}. Furthermore, we chose to analyze the aforementioned period because it was a period with several controversies and potential for incivility due to Linus Torvalds' temporary break from his maintainer role~\cite{linusleaving}, as well as when the code of conduct was established in the Linux kernel community.

The review emails of LKML are stored in different git repositories\footnote{\url{https://lore.kernel.org/lkml/_/text/help/}}. We first used git commands to extract the email content of each repository and collected a total of 406,719 review emails in the studied period. Since a discussion to review a patch is spread across a multitude of email replies, we then group individual emails by email threads using the Mailboxminer tool~\cite{bettenburg2009empirical}. As a result, we found 55,396 email threads.

\subsection{Identifying rejected patches}
\label{sec:identify-rejected-patches}

In this paper, we focus on characterizing incivility in code review discussions of rejected patches. \CHANGED{We chose to study rejected patches because (1) rejected patches represent a majority (about 66\%) of all patches submitted to LKML~\cite{jiang2013will} and (2) rejection indicates conflict, thus a greater potential for identifying and categorizing (in)civility. A previous work has found that even when patches were rejected for technical reasons, the language used can be harsh and toxic~\cite{alami2019does}.}

Thus, after collecting code review emails, it is necessary to identify which patches were rejected. While most open source projects use web-based review environments like Gerrit or GitHub's pull requests, the Linux kernel community's usage of regular mailing lists for review discussions implies that the review decision (accept/reject) of a given review email is either not explicitly recorded or mentioned in an inconsistent manner% ., and has to be reconstructed by trying to match   leads to the review discussions being disconnected from the review decision (accept/reject) are separated from the committed code on Git (as it would not be in modern code review environments such as Gerrit)
~\cite{jiang2014tracing}. Hence, to identify whether patches discussed in the review emails are eventually accepted or rejected, we need to link the code review emails to the commits in the git repository. If a patch is both in the mailing list and the git repository, then the patch was accepted. Otherwise, the patch was rejected.

To link review emails to commits, we developed three heuristics based on the work of Jiang et al.~\cite{jiang2013will,jiang2014tracing}. The heuristics determine that a patch mentioned in a review email is accepted (i.e., linked to a commit) if: (1) the subject of the email is the \textit{same subject} of the commit message, (2) the email and the commit are associated with the \textit{same author} and the commit appeared later than the email, or (3) the email includes code that changed the \textit{same file} as the commit and the file includes a considerable proportion (based on a threshold) of changed lines as the commit file. If one email in an email thread is determined as being linked to an accepted patch, all the emails in that thread are then considered related to the accepted patch.

We used the linux-tips mailing list~\cite{jiang2014tracing} to validate the performance of these heuristics, since this mailing list contains review emails listing, for each accepted patch, the identifier of the git commit that was generated for the patch (i.e., it contains a gold standard)% identifier indicates  contains not only emails with patches, but also indicates  information of the commits for accepted patches as a gold standard
. We calculate the \textit{precision} based on the ratio of equal links found by our heuristics and the gold standard to the number of links found by our heuristics. The \textit{recall} is the ratio of equal links found by our heuristics and the gold standard to the number of links found by the gold standard. We found that the three heuristics achieved a precision of 98.51\% and a recall of 90.08\% when identifying review emails that linked to git commits. After applying these heuristics to our dataset, we identified 26,989 (48.72\%) rejected threads.

\subsection{Filtering and sampling rejected email threads}

We automatically performed the following filtering steps on all identified rejected email threads from the previous step to remove email threads that did not contain a discussion or were in fact associated with an accepted patch (i.e., false positives of the heuristics). 

\begin{enumerate}

\item We excluded threads that do not include a patch (source code snippets) in the emails; that is, threads that had just discussions. We removed 11,074 threads in this step.

\item We excluded patch submission threads with only one email (i.e., without a followup by someone from the community). We removed 3,446 threads in this step.

\item We excluded threads with only one response in which the response has the following keywords: \textit{``Applied to''}, \textit{``Applied''}, \textit{``Queueing for''}, \textit{``Queued''}, \textit{``Tested-by''}, \textit{``Reviewed-by''} and \textit{``Acked-by''}. These keywords indicate that the patch might have been accepted, or there is no discussion that we can analyze in the context of this paper. We removed 887 threads in this step.

\end{enumerate}

After applying all the filters, 11,582 (42.91\%) rejected threads were retained. We then randomly sampled 372 threads, achieving a confidence level of 95\% under a confidence interval of 5. We performed a manual verification on these 372 threads and 110 of those that mentioned that the patch was accepted were removed from our analysis (i.e., false positives of the heuristics). The remaining 262 email threads comprised 1,545 code review discussion emails (i.e., emails that reply to the original submission of the patch), which were the focus of our qualitative analysis. 

% Applied to my 4.16 branch
% Applied to the spi-nor/next branch of linux-mtd
% Applied this one.
% All now queued up, thanks.
% I've applied v13 which indeed does have this change.
% Applied to the togreg branch of iio.git where it will sit until after the merge window.
% Applied, thanks.
% That patch will go to 4.16 and from there to 4.15 via the stable tree.
% I'll pick your patch into our kernel.
% and no discussion further

\subsection{Qualitative coding on 262 rejected email threads}
\label{sec:qualitative-analysis}

To answer our RQs, we did a qualitative analysis~\cite{strauss1987qualitative, strauss1990open} on the sample of 262 email threads that were composed of 1,545 code review discussion emails. The coding focused on the following aspects. %We did not include the submission of patches and RFC emails (usually represented by [PATCH] and [RFC] in the email subject) in our sample. In other words, we just included the replies to those emails. \\

\subsubsection{Identification of tone-bearing discussion features (TBDFs)} Recall that we define TBDF as \textit{conversational characteristics demonstrated in a written sentence that convey a mood or style of expression}. We inspired our work on the uncivil discussion features proposed by Coe et al.~\cite{coe2014online}, namely name-calling (mean words directed at a person or a group of people), aspersion (mean words directed at an idea, plan, or behavior), lying accusation (stating that an idea, plan, or policy was a lie), vulgarity (usage of profanity or not a proper language in a professional discourse), and pejorative for speech (criticizing the way a person communicates). The first author conducted an inductive coding~\cite{thomas2003general}, \CHANGED{which is an approach that allows the research findings to emerge from the interpretation of the raw data,} on each email of each sampled thread to manually identify TBDFs; the coding was conducted on the sentence-level within each email.

To identify TBDFs in relevant sentences of an email, we take into consideration the context of the previous emails in the thread. \CHANGED{We decided to analyze the previous context due to the insights gathered in our pilot study (see Section~\ref{sec:case-study}). Furthermore, sentences that contain purely technical discussions were not coded because they do not convey a mood or a style of expression. More specifically, we do not code if the sentence is:

\begin{itemize}
    \item discussing about the program's behavior, such as in ``\textit{Seems like kallsyms would be one to absolutely scan... it shouldn't cause hangs either.}''
    \item asking purely technical questions, especially if the previous email in the thread is also purely technical. For example, after someone has submitted a patch, the answer was ``\textit{With stock knob settings, that's too late to switch from llc -> l2 affinity for sync wakeups, and completely demolished tbench top end on huge socket NUMA box with lots of bandwidth.}''
    % \item simply making a suggestion about the patch, such as ``\textit{How about adding an additional patch on top taking into account the ignore\_children flag and folding that into the series?}''
    % \item simply making a request, such as ``\textit{Could you please just merge the fix instead?}'' or ``\textit{Could you please defer this to v4.17?}''
    \item an explanation without any mood or style of expression. For example, ``\textit{I already have an equivalent change queued up.}''
    % \item not expressing the speaker's own mood or style of expression, such as in ``\textit{This issue seems to be caused by patch [1/2].}''
\end{itemize}
} % For example, if we observe that people are arguing in the emails preceding the one that is being analyzed, we classified the TBDF of such email with a negative tone \bram{even if the email-under-analysis is positive? also, the whole email is tagged with one TBDF, or sentence-level?}. However, if we observe that people are having a respectful conversation, we classified the TBDF of such email with a positive tone. 
Codes related to the identified TBDFs were added in the codebook~\cite{saldana_coding} during the process with a definition and one or more examples. The codebook containing all manually identified TBDFs was iterated based on the discussion with two other authors. To refine the codebook and guarantee that the inductive coding can be replicated, \CHANGED{the second author deductively coded all emails in which the first author has identified TBDFs, adding up to 191 emails. The second author did not assess the emails in which the first author judged to contain only technical information, as they are easy to be classified without ambiguity with the criteria listed above.} Similar to the initial coding process, when identifying the TBDFs of a specific email, the second author was asked to read all the previous emails in the thread to understand the content. \CHANGED{We computed the Cohen's Kappa to evaluate inter-rater reliability of our coding schema~\cite{mcdonald2019reliability}. Results show that the Kappa scores on all codes in the final version of the codebook ranged from 0.42 to 0.96, with an average of 0.62, demonstrating a substantial agreement\CHANGEDFINALVERSION{~\cite{viera2005understanding}}. The complete codebook can be found in our online repository\footnote{\replicationPackage}.}  %We later discussed in detail and reached an agreement on the coding.
%The codebook was updated and used by the first author to adjust the coding for the undiscussed emails. 
The results of this step were used to answer \textit{RQ1: Which features of discussion can be found in code reviews of rejected patches?} and \textit{RQ2: How much incivility exists in code review discussions of rejected patches?}. \CHANGED{We answer RQ1 by describing the TBDFs found with the inductive coding, and we answer RQ2 by presenting the frequency of (in)civility as well as the frequency of each TBDF in the sentence, email, and thread levels.}

%two other researchers deductively coded 300 emails using the codebook~\cite{lombard2002content, distaso2012multi} based on a stratified sample of civil and uncivil emails identified by the first author. 

%We computed the inter-rater reliability using the Gwet's AC1 test~\cite{gwet2008computing}, and we found an agreement coefficient of \hl{X} with rater 1, and a coefficient of \hl{Y} with rater 2. \bram{why AC1?}

\subsubsection{Identification of email and thread attributes}

We also explore email and thread attributes that might be associated with the occurrence of incivility to answer \textit{RQ3: How is incivility correlated with the occurrence of arguments, the individual contributors, and the discussion topics?}

To assess if incivility is correlated with arguments, we coded for the occurrence of an argument in each email thread based on all email discussions in the thread. In our context, an argument happens if two parties (usually a developer and a maintainer) disagree with each other and both voice their opinions. \CHANGED{We analyzed our results by showing the frequency of threads and emails with and without an argument. %Since the distribution of our data is normal, 
We performed the t-test~\cite{kim2015t} to assess if there is a statistical difference between (i) the length of the discussion of threads with and without an argument, and (ii) the number of civil and uncivil emails in threads with and without an argument. Additionally, using the chi-square test~\cite{mchugh2013chi}, we assessed the relationship between (i) the occurrence of an uncivil email in a thread and (ii) the occurrence of an argument in a thread. Finally, we computed the effect size of the relationship between the two aforementioned variables using Cramer's V~\cite{mchugh2013chi}.}

To investigate if incivility is correlated with individual contributors, for each analyzed email, we classified if the email author is a developer or a maintainer. We considered all individuals whose name is listed in \CHANGED{the most recent version of} the \texttt{MAINTAINERS} file\footnote{\url{https://github.com/torvalds/linux/blob/master/MAINTAINERS}, \CHANGED{last access: 2021-04-06}}, \CHANGED{which is an official file that lists the Linux kernel's maintainers}, as maintainers and all other individuals as developers. 
%\CHANGED{Following the same approach as~\cite{amreen2020alfaa}, 
\CHANGED{Additionally, we cluster individuals that have either the same name or the same email address together to accurately determine developers' identities and to avoid classifying the same person more than once as a developer or a maintainer. Then, we manually checked if all identities were correctly clustered together. We analyzed our results by presenting (i) the number of contributors that have sent technical, civil, and uncivil emails, and (ii) the distribution of technical, civil, and uncivil emails sent by developers and maintainers. Then, we compared with the t-test~\cite{kim2015t} if there is a statistically significant difference between the number of civil and uncivil emails sent by someone that has sent at least one uncivil email.}

Finally, to evaluate if incivility is related to specific topics of the discussion, we inductively coded for the discussion topic of each uncivil email based on the email subject and content. For terms that we do not understand in the email subject or content, we searched and read related material to have a better understanding. We then grouped our codes into categories to topics. \CHANGED{We analyzed the frequency of the encountered categories of topics in emails sent by developers and maintainers.}

\subsubsection{Identification of discoursal causes of incivility.} To answer \textit{RQ4: What are the discoursal causes of incivility?}, we consider the content of the email that the uncivil email is replying to. This allows us to grasp the context and identify what triggered the incivility in practice. We then conducted \CHANGED{open coding~\cite{williams2019art}} on each sentence classified with an uncivil TBDF for their causes. During our coding, one sentence might have several causes. Finally, we did a \CHANGED{thematic analysis~\cite{williams2019art}} on the identified causes to group them into themes. \CHANGED{We answer RQ4 by describing the causes and the frequency of each cause in emails sent by developers and maintainers.}

\subsubsection{Identification of discoursal consequences of incivility.} To answer \textit{RQ5: What are the discoursal consequences of incivility?}, we followed a similar approach as \CHANGED{for the causes of incivility} but focused on the email that replied to the uncivil email. We conducted open coding on the consequences of each sentence classified with an uncivil TBDF and grouped the codes into themes. \CHANGED{We answer RQ5 by describing the consequences and the frequency of each consequence in emails sent by developers and maintainers. Additionally, we analyze the relationship between the causes (RQ4) and consequences (RQ5) of emails sent by developers and maintainers.}
\section{Results}

In this section, we present the results to answer our five research questions that aim to characterize \CHANGED{incivility} in code review discussions of rejected patches.

\subsection{RQ1. Tone-bearing discussion features (TBDFs) in code review discussions of rejected patches}
\label{sec:rq1-results}

To answer RQ1, we identified \CHANGED{16} TBDFs through the open coding of the code review discussions of rejected patches. We further grouped these TBDFs into positive, neutral, and negative features according to the general tone expressed. Finally, we separated uncivil features from the negative ones if the sentence includes a feature that conveys an unnecessarily disrespectful tone. Because of our focus on uncivil communication, we identified more fine-grained uncivil features. We describe these TBDFs in the following sections.

\subsubsection{Positive features}
\begin{itemize} [leftmargin=12px]
    
    \item \textbf{Appreciation \CHANGED{and excitement}.} Code review discussion participants have expressed appreciation, enthusiasm, and interest towards certain problems, solutions, or discoveries. For example, one participant was excited over a discovery of a technical approach: ``\textit{All this time, I thought these parameters were for power gating... I also did not expect that clock gating had to be disabled before we could program them. Great find!}''
    
    \item \CHANGED{\textbf{Considerateness.} This feature appears in sentences that express extreme polite requests made in form of questions or in expressions considerate of other people's opinions. For example, ``\textit{My point is, we might as well take the opportunity to fix this right away, don't you think?}''}
    
    \item \CHANGED{\textbf{Humility.} This feature appears when participants express in a modest way that they did not understand something, they need to ask for someone's opinion or help, and/or they recognize someone's efforts. ``\textit{The patch does more than described in the subject and commit message. At first I was confused why do you need to touch here. It took few minutes to figure it out.}''}
    
\end{itemize}

\subsubsection{Neutral features}
\begin{itemize} [leftmargin=12px]
    
    \item \CHANGED{\textbf{Friendly joke.} This appears when someone is making a suggestion or a statement in form of a joke. We consider this feature as neutral because expressions coded with this code are mostly used to address awkward or unpleasant situations. For example, ``\textit{Instead of hitting the fly, hit "make htmldocs" on the keyboard :)}'' and ``\textit{Do you believe me now, that [programming details] is not "the whole and only reason" I did this? :D}''}
    
    \item \CHANGED{\textbf{Hope to get feedback.} This appears when someone hopes/wishes to get feedback from the community or individuals that are more knowledgeable about a specific problem. For example ``\textit{It would be good to get comments from people more [programming details] knowledgeable, and especially from those involved in the decision to do separate [programming details].}''}

    \item \CHANGED{\textbf{Sincere apologies.} This code is used to capture expressions in which participants say sorry about what they did not do and/or because of a wrongdoing (e.g., someone is being harsh). For example, ``\textit{I am sorry that I didn't join the discussion for the previous version but time just didn't allow that. So sorry if I am repeating something already sorted out.}''}

\end{itemize}

\subsubsection{Negative features}
\begin{itemize} [leftmargin=12px]

    \item \CHANGED{\textbf{Commanding.} This feature appears in sentences that issue a command, instructions, or a request in an abrupt way. Someone might also ask rhetorical questions to express an order or command. For example, a maintainer asked a developer: ``\textit{Do not use attachments to fix this problem, the patch must be inline after your commit message and signoffs.}''}
    
    \item \textbf{Oppression.} Code review discussion participants, especially developers, sometimes expressed resistance or reluctance when forced to adopt a solution or an approach by a person of power (e.g. a maintainer). Furthermore, they might express mental pressure or distress. For example, ``\textit{When one of the authors of the original document objected, I felt it is better to backoff. But if there is a consensus, I will proceed.}''
    
    \item \CHANGED{\textbf{Sadness.} This feature appears when the speaker is unhappy or sorrowful because the result was not as expected. Additionally, someone of less power (usually a developer) might experience a condition put to them that negatively affects their feelings. For example, ``\textit{I'll remember all this for the next time (if next time there is, of course, I was already quite hesitant to spend time to prepare and send patches for these issues with [programming details] mix-up).}''}

\end{itemize}

\subsubsection{Uncivil features}
\begin{itemize} [leftmargin=12px]

    \item \textbf{Bitter frustration.} This feature appears when someone expresses strong frustration when addressing a false accusation or a lie, expressing that expectations are not met, voicing dissatisfaction or annoyance due to a lack of information or explanation, dealing with erroneous assumptions, or describing a problem that was not mentioned before. For example, when reviewing a piece of code that was submitted as rich text in an email rather than as plain ASCII text, a maintainer wrote: ``\textit{I cannot apply a patch which has been corrupted by your email client like this.}''
    
    \item \CHANGED{\textbf{Impatience.} Participants might demonstrate impatience when they express a feeling that it is taking too long to solve a problem, understand a solution, or answer a question. Furthermore, impatience appears when someone has to repeat the same information over and over again, someone is doing a repeated mistake, and/or not everyone is participating in the discussion. For example, ``\textit{Note instead the time lapse between this and previous posting of the series, and if you want to assume something, assume things can get missed and forgotten without intent or malice.}''}

     \item \textbf{Irony.} In a few cases, contributors used expressions that usually signify the opposite in a mocking or blaming tone. For example, one contributor wrote on a late response to a maintainer's comments: ``\textit{Only about a year and a half late, nice!''}
     
     \item \CHANGED{\textbf{Mocking.} This feature appears when a discussion participant is making fun of someone else, usually because that person has made a mistake. For example, ``\textit{I would also suggest that your time might be spent more productively if you would work on some more useful projects. There is more than enough to do. However, that's up to you.}''}
    
    \item \textbf{Name calling.} This appears in sentences that include mean or offensive words directed at a person or a group of people. \CHANGED{For example, ``\textit{If you want to provide more accurate documentation then you better come up with something which is helpful instead of a completely useless blurb like the below...}''}
    
    \item \CHANGED{\textbf{Threat.} In a few cases, contributors put a condition impacting the result of another discussion participant or that person's career. For example, ``\textit{Unless you have solid suggestions on how to deal with all of them, this is a complete non-starter.}''}
    
    \item \textbf{Vulgarity.} In some cases, contributors used profanity or language that is not considered proper in professional discourse.
    
\end{itemize}
\subsection{RQ2. Frequency of incivility in code review discussions of rejected patches}
\label{sec:frequency-incivility}

To understand the frequency of incivility and answer RQ2, we classified code review emails into the following three categories based on the TBDF they demonstrate (found in RQ1).

\begin{itemize}
    \item \textbf{Uncivil:} An email is classified as \textit{uncivil} if it has at least one sentence demonstrating an uncivil TBDF (i.e., \textit{bitter frustration}, \CHANGED{\textit{impatience}}, \textit{irony}, \CHANGED{\textit{mocking}}, \textit{name calling}, \CHANGED{\textit{threat}}, or \textit{vulgarity}).
    
    \item \textbf{Civil:} An email is classified as \textit{civil} if it has at least one sentence labeled with a TBDF, but none of the TBDF is uncivil. 
    
    \item \textbf{Technical:} An email is classified as \textit{technical} if it has no sentence labeled with a TBDF. This indicates that the discussion is focused only on technical aspects and does not include any perceivable emotion or tone.
\end{itemize}

\textbf{We found that \CHANGED{1,377 (89.13\%)} of the manually analyzed emails are technical (i.e., does not contain a TBDF), \CHANGED{112 (7.25\%)} are uncivil, and \CHANGED{56 (3.62\%)} are civil.} Because of the technical focus of the LKML, it is natural that most of the emails are technical-oriented and did not bear any TBDF. However, to our surprise, more than half of the non-technical emails are uncivil (\CHANGED{66.66\%} of non-technical emails). On average, email threads with at least one uncivil email included \CHANGED{1.96} uncivil emails (ranged from 1 to \CHANGED{11}).

% 56	civil
% 51	review
% 1326	technical
% 112	uncivil
% 1545 TOTAL
% 168 NON-TECHNICAL EMAILS

Interestingly, \CHANGED{27} of the \CHANGED{112} uncivil emails \CHANGED{(24.11\%)} also included civil discussion features. Specifically, \CHANGED{\textit{humility} appeared in nine uncivil emails, \textit{commanding} appeared in eight uncivil emails, \textit{sadness} in seven emails, \textit{hope to get feedback} appeared in five, \textit{considerateness} appeared in three uncivil emails, \textit{sincere apologies}, \textit{oppression}, \textit{friendly joke} and \textit{appreciation and excitement} appeared in one uncivil email.} This result indicates that uncivil comments can sometimes contaminate a discourse. 

% 9	humility
% 8	commanding
% 7	sadness
% 5	hope_to_get_feedback
% 3	considerateness
% 1	sincere_apologies
% 1	oppression
% 1	friendly_joke
% 1	appreciation_excitement

\begin{figure}[t]
\centering
\includegraphics[clip, width=\linewidth]{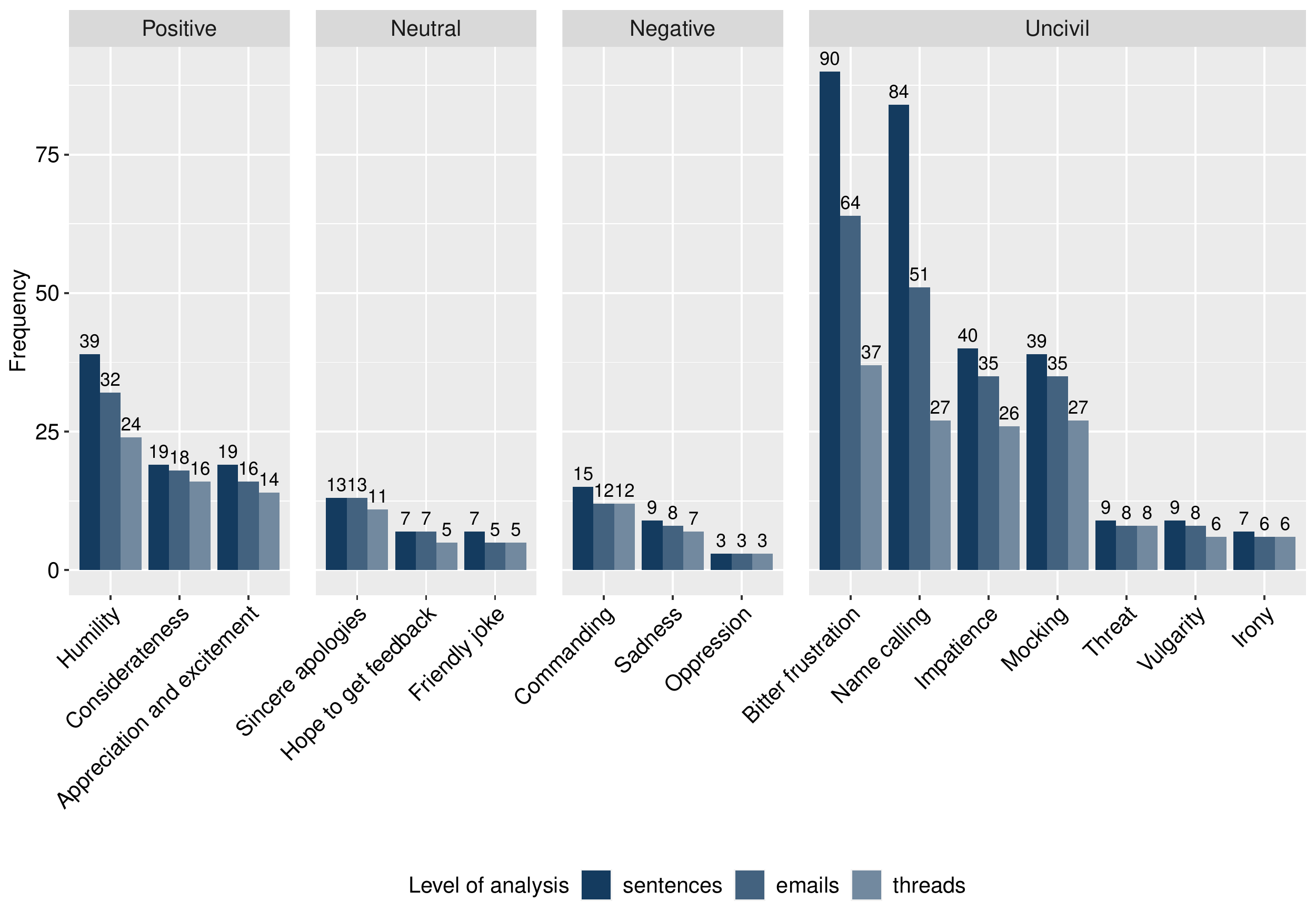}
\caption{\CHANGED{Frequency of TBDFs in code review discussions of rejected patches. \textit{Note:} A sentence can be coded with multiple codes.} 
}
\label{fig:frequency-features-discussion}
\end{figure}

Figure~\ref{fig:frequency-features-discussion} summarizes the frequency of each TBDF in the sentence, email, and email thread levels. \CHANGED{We observe that \textit{humility} is the most frequent feature for positive TBDFs (39 sentences out of 32 distinct emails). Although \textit{commanding}, \textit{sadness} and \textit{oppression} convey a negative tone, they do not happen very often (only 15, 9, and 3 sentences, respectively). Finally, \textit{bitter frustration}, \textit{name calling}, \textit{impatience}, and \textit{mocking} are the most frequent uncivil TBDFs. Interestingly, these results match the ones found in our motivational case study (see Section~\ref{sec:case-study}), where participants mentioned that civility is related to \textit{being humble} (most frequent positive TBDF), and that \textit{frustration} (most frequent uncivil TBDF) is a factor that can make communication uncivil.}

\subsection{RQ3. Correlations of incivility with email and thread attributes}

In order to explore factors that might explain the appearance of incivility (RQ3), we analyzed the correlations of uncivil communication with three email and thread attributes: the occurrence of an argument in the thread, the author of uncivil emails, and the topic discussed in the thread. These correlations are the so-called devil's advocate arguments since they might provide the most obvious explanations for uncivil communication during code review.% someone would except that the aforementioned factors would lead to incivility.

\subsubsection{Correlation of incivility with argument in the thread} 

Previous work~\cite{angouri2012theorising} has found that arguments are typically related to confrontation and conflicts, and have, consequently, negative effects. We define the appearance of an \textit{argument} in a code review email thread as two parties (usually a developer and a maintainer) disagreeing with each other and each laying out their reasons (see Section~\ref{sec:qualitative-analysis}). Based on that, RQ3 hypothesizes that incivility is correlated with arguments.

Using the above definition for ``argument'' to code the email threads, \textbf{we identified that only 10.31\% of the email threads in our dataset included an argument.} They cover 23.37\% (361) of the emails in our sample, among which \CHANGED{77.56\%~(280)} were technical emails, \CHANGED{5.82\%~(21)} were civil emails and \CHANGED{16.62\%~(60)} were uncivil emails (see Table~\ref{tab:frequency-disagreement}).

Conversely, among all uncivil emails in our dataset, \CHANGED{53.57\%} were part of a thread with argument; this percentage was \CHANGED{37.50\%} for civil emails and \CHANGED{20.33\%} for technical emails. While these results show that more than half of the uncivil emails were indeed related to the presence of an argument in a thread, they also imply that, \textbf{against our expectations, almost half of the uncivil emails \CHANGED{(46.43\%)} were part of threads \emph{without} an argument.}

\begin{table}[ht]
\centering
\small
\aboverulesep=0ex
\belowrulesep=0ex
\caption{\CHANGED{Frequency of threads and emails with or without an argument in code review discussions of rejected patches.}}
\label{tab:frequency-disagreement}
\begin{tabular}{l|c|ccc|c}
\toprule
\multicolumn{1}{c|}{\textbf{Thread code}} & \textbf{\begin{tabular}[c|]{@{}c@{}}\#email\\ threads\end{tabular}} & \textbf{\begin{tabular}[c]{@{}c@{}}\#technical\\ emails\end{tabular}} & \textbf{\begin{tabular}[c]{@{}c@{}}\#uncivil\\ emails\end{tabular}} & \textbf{\begin{tabular}[c]{@{}c@{}}\#civil\\ emails\end{tabular}} &  \textbf{\begin{tabular}[c]{@{}c@{}}Total \\ emails\end{tabular}} \\ \midrule
\textbf{Without argument}  & 235    & 1097    & 52 & 35 & 1184 \\
\textbf{With argument}  & 27  & 280  & 60 & 21 & 361\\ \midrule
\textbf{TOTAL} & 262 & 1377 & 112 & 56 & 1545    \\ \bottomrule                                
\end{tabular}
\end{table}

\begin{figure}[t]
\centering
\includegraphics[clip, width=\linewidth]{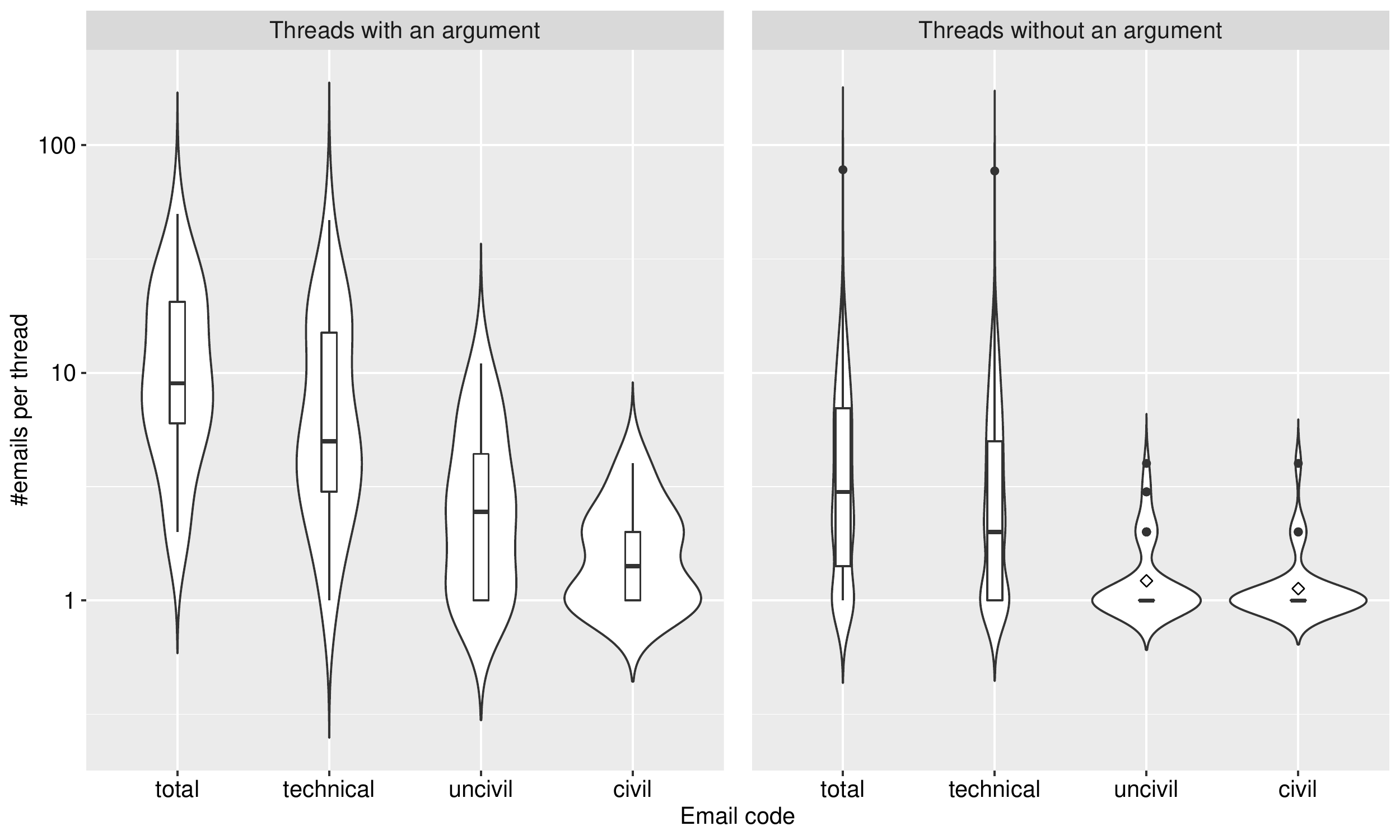}
\caption{\CHANGED{Distribution of number of emails for each email type in threads with or without an argument.}}
\label{fig:distribution-email-agreement-disagreement} 
\end{figure}

Figure~\ref{fig:distribution-email-agreement-disagreement} presents the distributions of the number of emails for each email type (i.e., technical, uncivil, or civil) in a thread with or without an argument. We observe that on average threads with an argument tend to have longer discussions (\CHANGED{13.37} emails) than threads without an argument (\CHANGED{5.04} emails); a t-test indicated that this difference is significant (\CHANGED{$t=3.75, p=0.0008$}). Among the threads that contain an argument, the difference between the average number of uncivil emails \CHANGED{(3.33) and that of civil emails (1.75) is also significantly different ($t=-2.21, p=0.04$).} This difference is not statistically significant in threads without an argument (\CHANGED{$t=-0.78, p=0.43$}), which on average contains \CHANGED{1.21} civil emails and \CHANGED{1.33} uncivil emails. A chi-square test indicated that there is a significant relationship between (1) the occurrence of uncivil emails in a thread and (2) the occurrence of an argument in a thread%. Threads with an argument are more likely to have uncivil emails
\CHANGED{ ($X^2 (1, N = 1545) = 12.99, p=0.0003$)}. However, the association between these two variables is very weak, with Cramer's \CHANGED{$V=0.19$}.

% length - argument total
% > summary(total_argument)
%   Min. 1st Qu.  Median    Mean 3rd Qu.    Max. 
%   2.00    6.00    9.00   13.37   20.50   50.00 
% length - no argument total
% > summary(total_no_argument)
%   Min. 1st Qu.  Median    Mean 3rd Qu.    Max. 
%   1.000   1.500   3.000   5.038   7.000  78.000 
  
% > summary(civil_argument)
%   Min. 1st Qu.  Median    Mean 3rd Qu.    Max. 
%   1.00    1.00    1.50    1.75    2.00    4.00 
% > summary(uncivil_argument)
%   Min. 1st Qu.  Median    Mean 3rd Qu.    Max. 
%   1.000   1.000   2.500   3.333   4.500  11.000 
  
% > summary(civil_no_argument)
%   Min. 1st Qu.  Median    Mean 3rd Qu.    Max. 
%   1.000   1.000   1.000   1.207   1.000   4.000 
% > summary(uncivil_no_argument)
%   Min. 1st Qu.  Median    Mean 3rd Qu.    Max. 
%   1.000   1.000   1.000   1.333   1.000   4.000 
  
\subsubsection{Correlation of incivility with email authors}

Liu et. al.~\cite{liu2009explaining} found that some factors related to personal characteristics are positively correlated to incivility. Based on that, we hypothesized that incivility is correlated with the authors' personality or the overall email tone.

\begin{figure*}[t] 
\subfloat{\includegraphics[clip, width=0.3\linewidth]{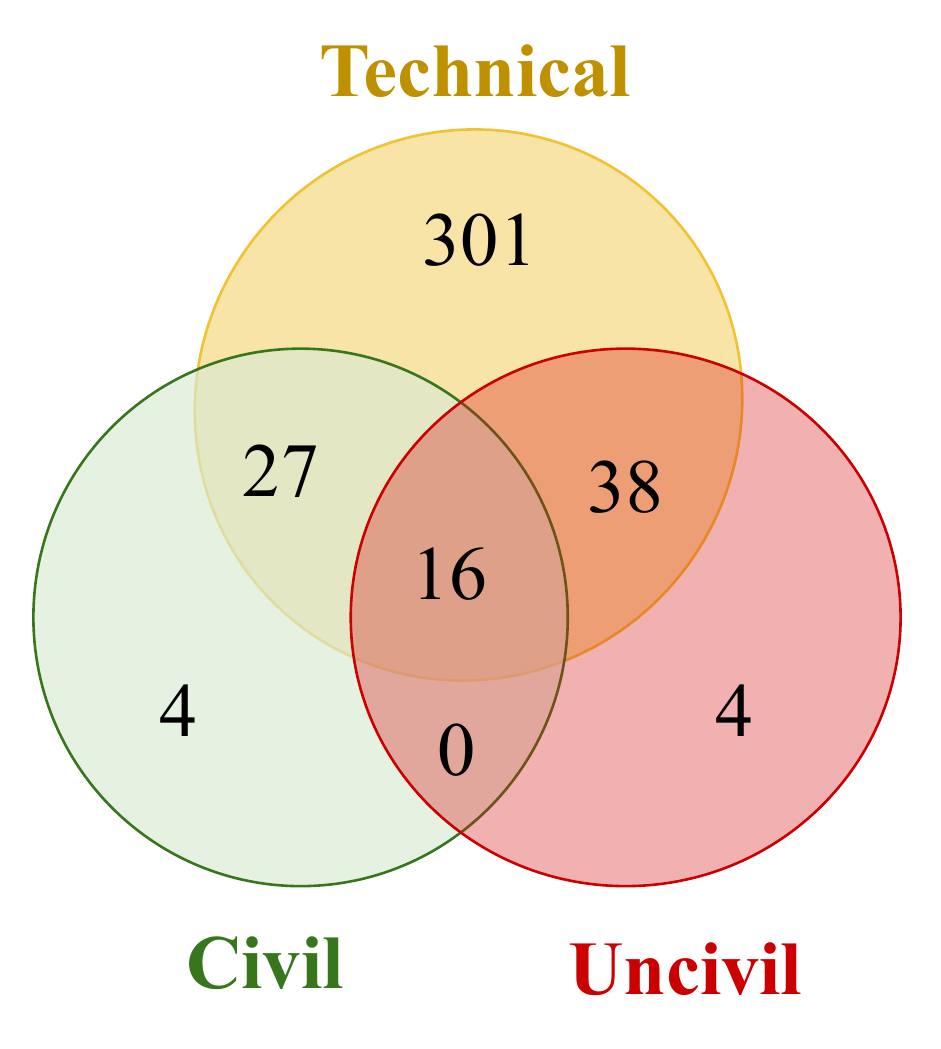}}
\caption{\CHANGED{Number of email authors per email type.}}
\label{fig:authors_per_email_code} 
\end{figure*}

Among all the authors of emails in our dataset (\CHANGED{$N=390$}), most (\CHANGED{382 or 97.94\%}) have sent at least one technical email, \CHANGED{47 (12.05\%)} have sent at least one email classified as civil, and \textbf{\CHANGED{58 (14.87\%)} have sent at least one uncivil email}. Within the last group, there were \textbf{\CHANGED{four} individuals \CHANGED{(two developers and two maintainers}) who have only sent uncivil emails}, while \CHANGED{54 people (14 developers and 40 maintainers)} have also sent civil and technical emails. Figure~\ref{fig:authors_per_email_code} presents the number of individuals that have authored each type of email.

%only_uncivil:  {'190', '196', '163', '338'}
% maintainer, developer, maintainer, developer

\begin{figure}[t]
\centering
\includegraphics[clip, width=\linewidth]{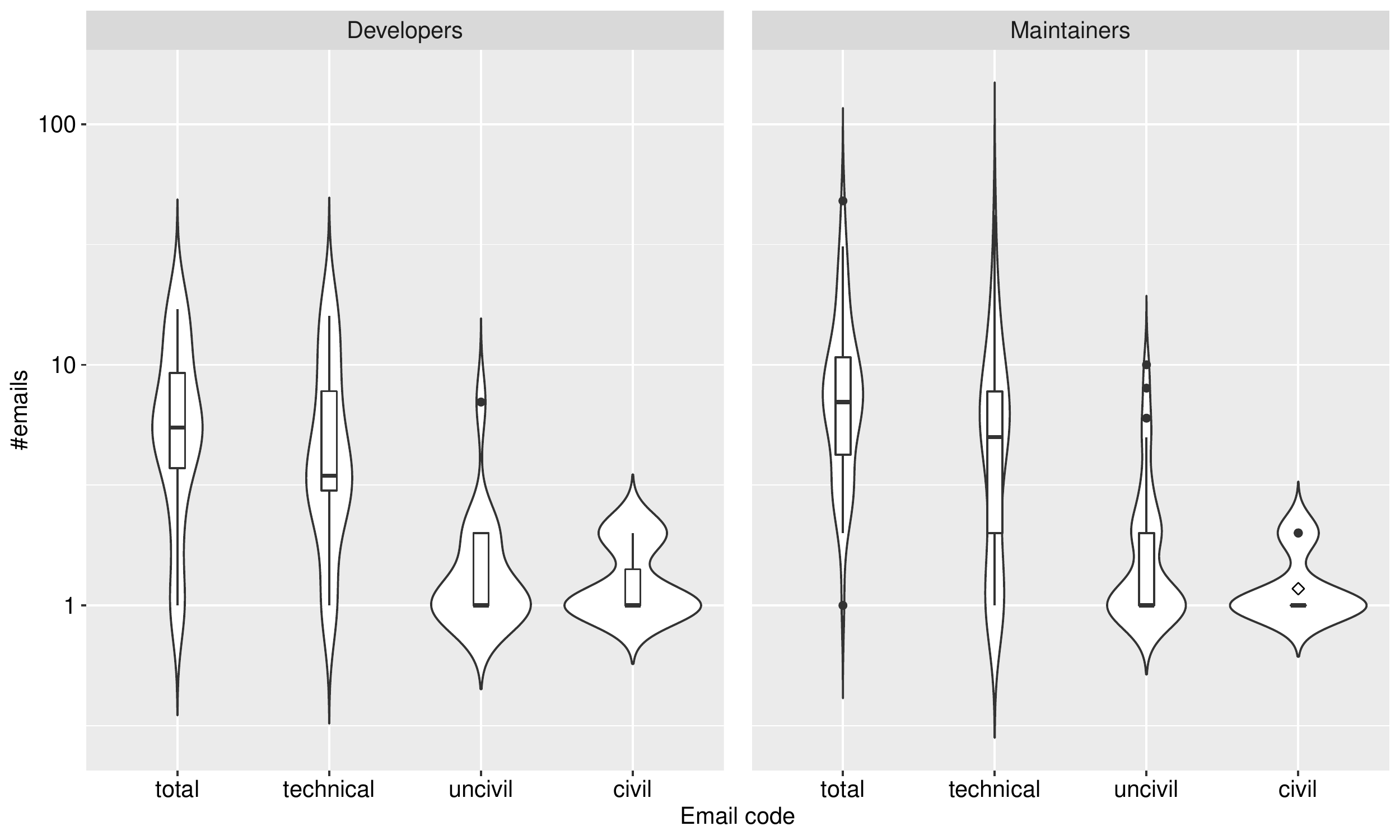}
\caption{\CHANGED{Distribution of emails sent by individuals that have sent at least one uncivil email.}}
\label{fig:distribution-uncivil-role} 
\end{figure}

When analyzing emails sent by authors who have sent at least one uncivil email, we found that developers have sent a total of \CHANGED{26 (23.21\%)} uncivil emails, and maintainers have sent a total of \CHANGED{86 (76.79\%)} uncivil emails. Figure~\ref{fig:distribution-uncivil-role} presents the distributions of the number of civil, uncivil, and technical emails authored by developers and maintainers who have sent at least one uncivil email. Developers have sent on average \CHANGED{1.62} uncivil emails (\CHANGED{$SD=1.50$}), \CHANGED{1.33} civil emails (\CHANGED{$SD=0.58$}), and \CHANGED{5.71} technical emails (\CHANGED{$SD=4.99$}). Maintainers have sent on average \CHANGED{2.05} uncivil emails (\CHANGED{$SD=1.98$}), \CHANGED{1.23} civil emails \CHANGED{($SD=0.44$}), and \CHANGED{6.59} technical emails (\CHANGED{$SD=7.35$}).

As a result, we observe that \CHANGED{there is a difference in the number of civil and uncivil emails sent by someone who has sent at least one uncivil email ($t=-2.53, p=0.014$)}. However, even though maintainers have collectively sent more uncivil emails, there is no statistically significant difference in the number of uncivil emails between individual developers and individual maintainers (\CHANGED{$t=-0.87, p=0.39$}).

\subsubsection{Correlation of incivility with the topic under discussion}

Coe et al.~\cite{coe2014online} found that incivility is often associated with several key contextual factors, including the topic of the discussion. To understand how discussion topics are associated with incivility in code review, we identified themes of discussion topics in uncivil emails (see Section~\ref{sec:qualitative-analysis}).

\textit{Topics associated with uncivil emails written by developers.} We found six main categories of topics that developers were uncivil about. In \CHANGED{ten} uncivil emails posted by developers, the main discussion topic itself was the \textbf{workflow}, in aspects such as documentation, development conventions, or contribution process. In \CHANGED{six emails, developers were uncivil about \textbf{system components}, such as the display and media components. In four emails, developers were uncivil on the topic of \textbf{technical implementation} of various system aspects, such as problems with coscheduling different processes, implementing optimization techniques, collecting kernel debugging and performance information, and synchronization issues. In two emails, developers were uncivil about \textbf{network}, such as when discussing about network protocols.} In \CHANGED{two emails}, developers were uncivil in emails about issues with \textbf{memory}, such as allocating memory and shared-memory variables. Finally, in \CHANGED{two} emails, developers were uncivil \CHANGED{when discussing about \textbf{cryptography} of asynchronous messages.}

\textit{Topics associated with uncivil emails written by maintainers.} 
We found six categories that maintainers were uncivil about. \CHANGED{Maintainers were mainly uncivil when discussing about \textbf{system components} (in 34 emails), such as the file system, drivers, controllers, and hardware interfaces. Additionally, maintainers were frequently uncivil about the topic of \textbf{workflow} (in 23 emails). In nine emails, maintainers were uncivil about \textbf{network}, such as the speed of ethernet devices and network protocols. Maintainers were also uncivil in \CHANGED{eight} emails discussing about \textbf{memory}, such as problems with allocating memory and shared-memory variables. \textbf{Technical implementation} was the topic of six emails in which the maintainers were uncivil, containing discussions about handling exceptions, assigning values to boolean variables, data race, duplicated headers, and runtime problems. Finally, in six emails, maintainers were uncivil about \textbf{cryptography} when discussing cryptography of asynchronous messages.}

Our results show that developers and maintainers are uncivil about the same topics, in which developers are mostly uncivil about workflow and maintainers about system components. However, we could not find any pattern to conclude that the topic of the discussion is correlated with incivility. Therefore, incivility can happen when discussing any topic.

\subsection{RQ4. Discoursal causes of incivility}
\label{sec:causes}
We have found in RQ3 that incivility \CHANGEDFINALVERSION{in code review discussions of rejected patches} is not strongly correlated to the most obvious explanations, i.e., arguments, individuals, or topics. Thus, the goal of this section is to analyze in more detail the causes of incivility \CHANGEDFINALVERSION{in such discussions}. In order to answer RQ4 and understand the immediate discoursal causes of uncivil TBDF for developers and maintainers, we coded these causes based on the email that the uncivil email replied to (see Section~\ref{sec:qualitative-analysis}). Through this analysis, we identified themes that caused developers' and maintainers' uncivil communication \CHANGEDFINALVERSION{in rejected patches}. %We present these themes in the following sections.

\subsubsection{Causes of incivility in developers' emails}

\begin{figure}[t]
\centering
\includegraphics[clip, width=0.85\linewidth]{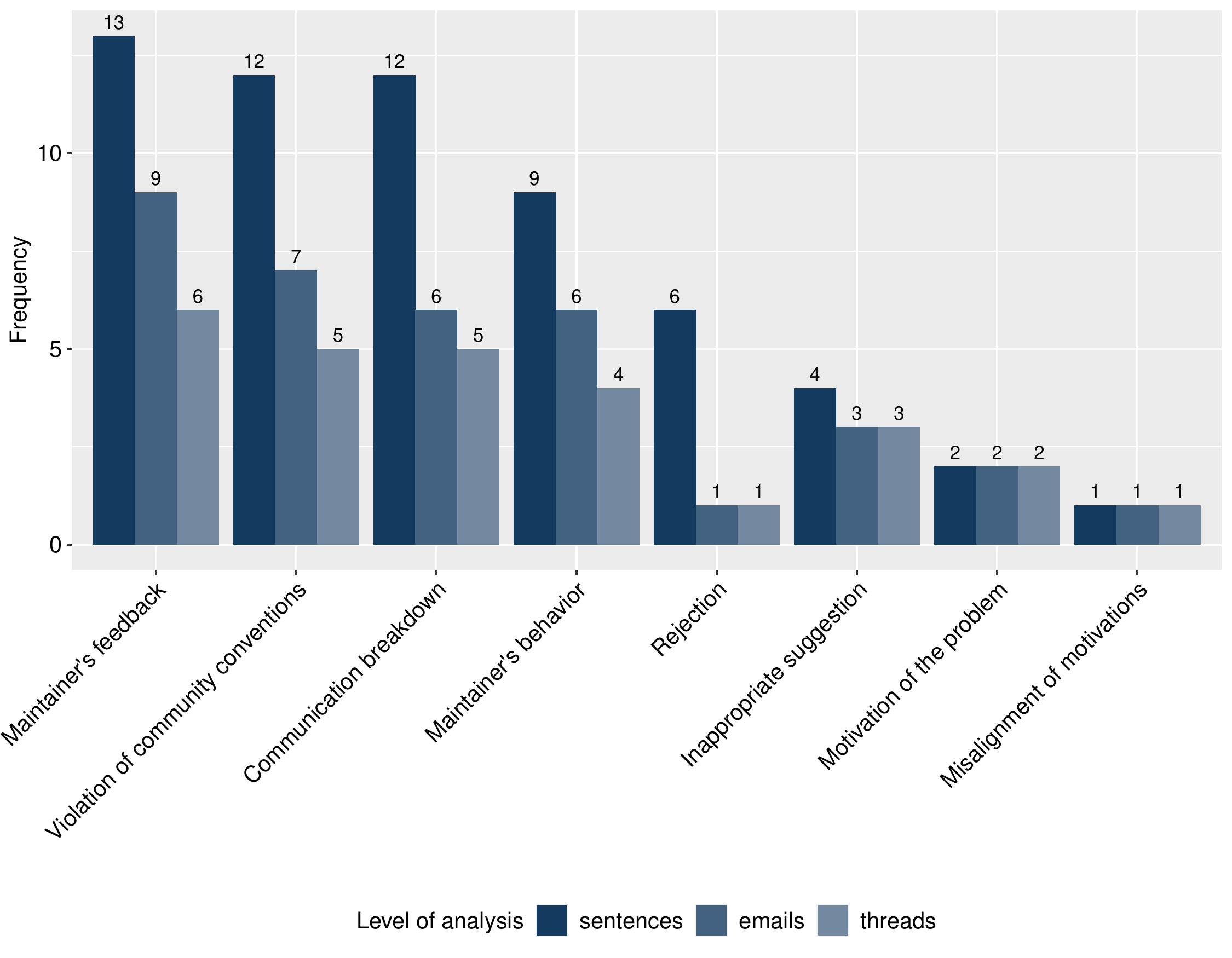}
\caption{\CHANGED{Frequency of causes of incivility in emails discussing rejected patches sent by developers. \textit{Note:} A sentence can be coded with multiple codes.} 
}
\label{fig:causes-developers}
\end{figure}

In total, we identified \CHANGED{eight} themes in the causes of incivility in the developers' emails (Figure~\ref{fig:causes-developers}). The most common categories are \CHANGED{the \textit{maintainer's feedback} (13 sentences), \textit{violation of community conventions} (12 sentences), and \textit{communication breakdown} (12 sentences).}

\textbf{Maintainer's feedback.} Some developers have been irritated by the maintainer's feedback. In most cases, the developer believed that the feedback proposed a non-optimal solution or a solution that can have a bad impact. For example, one developer reacted to a feedback suggesting that it is not the right time to fix the issue, writing: ``\textit{I don't think ``not fixing it because it's not fixed yet'' is a good reason to keep things the way they are.}''. Sometimes the developer also got frustrated because the maintainer asked to change direction or rejected the patch after a devoted effort from the developer.

\textbf{Violation of community conventions.} Some developers made uncivil comments due to disagreement with the workflow imposed by the community, not understanding the rationale behind a tedious workflow, or out of surprise by a workflow that they were not informed of. For example, disagreeing on the necessity of patch squashing (i.e., merging several commits into one), one developer wrote in an uncivil email: ``\textit{If you would insist on patch squashing, would you dare to use a development tool like ``quilt fold'' also on your own once more?}'' In another example, one developer did not know that an item in a workflow is necessary: ``\textit{Since when is the cover letter mandatory? ... for this simple test case addition what's the point?}''

\textbf{Communication breakdown.} Developers' uncivil comments were sometimes triggered by being misinterpreted by the maintainer or being unable to follow the maintainer's instructions. For example, a developer reacts to the maintainers' accusations, writing ``\textit{Wrong attitude what? I was trying to guess your reasoning ... since it wasn't clear to me why is your position what it is.}''. As an example of not being able to understand the maintainer, a developer wrote: ``\textit{I cannot comment on your proposal because I do not know where to find the reference you made.}'' 

\textbf{Maintainer's behavior.} Some developers' uncivil comments were direct results from a maintainer's uncivil behavior. In those cases, the developers tried to call out the uncivil behavior or to ask someone else to review the patch, nonetheless, in an uncivil way. For example, in a frustrated tone, a developer wrote: ``\textit{Would you like to answer my still remaining questions in any more constructive ways?}''

\textbf{Rejection.} Developers sometimes expressed frustration when they received a quick rejection or a rejection without sufficient explanation for the patches they submitted. For example, a developer complained, targeting a maintainer: ``\textit{I find it very surprising that you rejected 146 useful update suggestions so easily.}''

\CHANGED{\textbf{Inappropriate suggestion.} In a few cases, developers' uncivil comments were triggered when maintainers made an inappropriate suggestion. For example, a maintainer suggested a way of loading drivers instead of doing a mass code duplication. Then, the developer answered in a frustrated way: ``\textit{One do not load all [driver's name] at once, simply because one board has only one [driver's name] (or few closely related), and if one even try, almost none of them will initialize on given hardware.}''}

\textbf{Motivation of the problem.} A few uncivil comments made by developers originated from an argument with the maintainer on the relevance or importance of the problem. For example, in response to a maintainer that believed the developer was changing the symptom rather than the cause of the problem, a developer wrote, in frustration: ``\textit{It is not my theory guessing, it is a real problem..}''

\CHANGED{\textbf{Misalignment of motivations.} In one case, a developer and a maintainer had different opinions about the need of solving a specific problem. For example, the maintainer mentioned that the change proposed by the developer is not welcomed, and the developer wrote ``\textit{I think that's a pity.}''}

% \begin{table}[t]
% \small
% \caption{\CHANGED{Frequency of causes of incivility in emails \CHANGEDFINALVERSION{discussing rejected patches} sent by developers.}}
% \label{tab:causes-developers}
% \begin{tabular}{lccc}
% \toprule
% \textbf{Categories}           & \textbf{\#sentences*} & \textbf{\#emails} & \textbf{\#threads} \\ \midrule
% Maintainer's feedback      & 13     & 9        & 6                   \\
% Violation of community conventions  & 12    & 7 & 5                   \\
% Communication breakdown   & 12        & 6   & 5                   \\
% Maintainer's behavior       & 9       & 6    & 4                   \\
% Rejection        & 6       & 1         & 1                   \\
% Inappropriate suggestion & 4 & 3 & 3 \\
% Motivation of the problem & 2    & 2   & 2      \\ 
% Misalignment of motivations & 1 & 1 & 1 \\\midrule
% \textbf{TOTAL}            & \textbf{59}            & \textbf{35}       & \textbf{27}        \\ \bottomrule
% \end{tabular}
% \\\vspace{3pt}
% * A sentence can be coded with multiple codes.
% \end{table}
% we might have repeated quotes if the quote has more than one code

\subsubsection{Causes of incivility in maintainers' emails}

\begin{figure}[t]
\centering
\includegraphics[clip, width=0.85\linewidth]{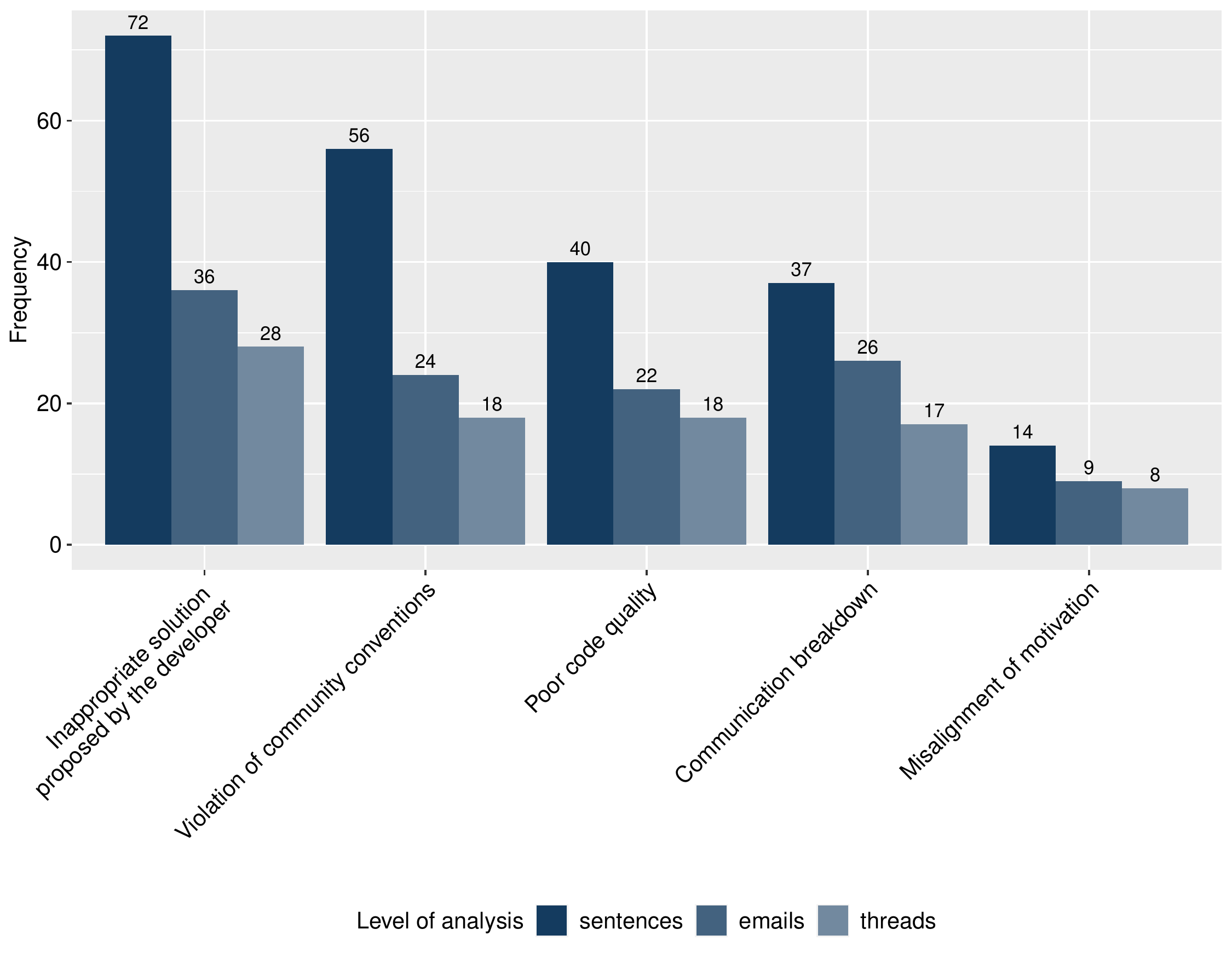}
\caption{\CHANGED{Frequency of causes of incivility in emails discussing rejected patches sent by maintainers. \textit{Note:} A sentence can be coded with multiple codes.} 
}
\label{fig:causes-maintainers}
\end{figure}

Concerning the main causes of incivility in emails sent by maintainers, we found five themes (summarized in Figure~\ref{fig:causes-maintainers}). The most common triggers for maintainers' uncivil comments are \CHANGED{\textit{inappropriate solution proposed by developer} (72 sentences), \textit{violation of community conventions} (56 sentences), and \textit{poor code quality} (40 sentences).}

\textbf{Inappropriate solution proposed by the developer.} Maintainers were most likely to get frustrated by the problems in the solutions proposed by the developers. This is sometimes because the proposed solution does not solve the problem; e.g., one maintainer told a developer: ``\textit{All in all, I'm not inclined to consider this approach, it complicates an already overly complicated thing and has a ton of unresolved issues while at the same time it doesn't (and cannot) meet the goal it was made for.}'' The maintainer could also get upset because the developer neglected important negative side effects or impacts when proposing their solution (e.g., one maintainer wrote, in frustration, ``\textit{You can't do it simply as it will cause deadlock due to nested locking of the buf\_lock.}'') or made uninformed changes to the existing code (e.g., \textit{``You are trying to "out smart" the kernel by getting rid of a warning message that was explicitly put there for you to do something.''}). Also, related to a communication issue, the maintainers were sometimes frustrated because they could not be convinced that the solution is valid; e.g., a maintainer commented on a patch: ``\textit{This looks really nonsensical and the commit message doesn't explain the rationale for that at all.}''

\textbf{Violation of community conventions.} Similar to developers, maintainers' uncivil comments were also triggered by a violation of community conventions related issues. This category included situations when the developers did not follow the workflow or they are not aware of certain steps in the workflow; e.g., ``\textit{The way you post them (one fix per file) is really annoying and takes us too much time to review.}'' Sometimes maintainers reacted in an uncivil way when the developer did not include sufficiently detailed commit messages, sent the patch to a wrong mailing list, did not follow the emailing convention, forgot to put the appropriate person in cc, etc.; e.g., ``\textit{Your email client should not be forcing you to top post. So please don't.}'' 

\textbf{Poor code quality.} Maintainers have also been annoyed by the quality of the code that the developers submitted. Often times, it is simply because the developer's code violates some best practices or conventions of programming or their code has readability issues; e.g., ``\textit{You implemented the same code thrice, it surely is not reduced.}'' Other times, it is because the developer's code is buggy and have run-time issues such as low performance; e.g., ``\textit{Did you actually test this?}''

\textbf{Communication breakdown.} Maintainers have made uncivil comments due to communication issues with the developer. In this category, the maintainers most frequently got frustrated due to insufficient explanation provided by the developer in the commit message or in the email discussion about the proposed patch. In such situations, the maintainers sometimes asked for explanations in a confrontational way; e.g., ``\textit{Now explain to me how you're going to gang-schedule a VM ... without it turning into a massive train wreck?}'' Sometimes, maintainers also got irritated because their comments seemed to be misunderstood or ignored by the developer; e.g. ``\textit{I think you didn't read my reply carefully. ... I'm just saying that the way you [solved this problem] is not at all the same as you would do [in another context]. Do you deny that?}''

\textbf{Misalignment of motivation.} Maintainers also demonstrated uncivil behavior when they disagreed with the developers' motivation to solve the problem; e.g. ``\textit{That is the whole and only reason you did this; and it doesn't even begin to cover the requirements for it.}'' Maintainers sometimes believed that the submitted patch is irrelevant or not useful, which triggered uncivil comments; e.g., ``\textit{Who the hell cares [about a technical solution]}'' The discussions of such issues are often about the problem space. maintainers sometimes believed that the developer did not realize the complexity of the problem or simply did not understand the problem; e.g., ``\textit{Either it does exist, or it doesn't. If it exists, it needs to be fixed. If it doesn't exist, nothing needs to be done. Which is the case?}''

% \begin{table}[t]
% \small
% \caption{\CHANGED{Frequency of causes of incivility in emails \CHANGEDFINALVERSION{discussing rejected patches} sent by maintainers.}}
% \label{tab:causes-maintainers}
% \begin{tabular}{lccc}
% \toprule
% \textbf{Category}           & \textbf{\#sentences*} & \textbf{\#emails} & \textbf{\#threads} \\ \midrule
% Inappropriate solution proposed by the developer & 72   & 36 & 28                \\
% Violation of community conventions     & 56     & 24    & 18                \\ 
% Poor code quality  & 40 & 22    & 18                                     \\
% Communication breakdown   & 37  & 26 & 17                               \\
% Misalignment of motivation  & 14    & 9    & 8                        \\
% \midrule
% \textbf{TOTAL}  & \textbf{219}  & \textbf{117}  & \textbf{89}     \\ \bottomrule
% \end{tabular}
% \\\vspace{3pt}
% * A sentence can be coded with multiple codes.
% \end{table}

\subsection{RQ5. Discoursal consequences of incivility}

After analyzing the causes of incivility (RQ4), we are interested in analyzing the consequences of incivility \CHANGEDFINALVERSION{in code review discussions of rejected patches} (RQ5). For that, we analyzed the next email that replied to the uncivil email (see Section~\ref{sec:qualitative-analysis}). Similar to the cause analysis of RQ4, we also evaluate the impact of uncivil emails separately for developers and maintainers.

\subsubsection{Consequences of incivility in developers' emails}
When developers are uncivil, we found \CHANGED{eight} main categories of consequences. In six emails, the maintainer stopped the review and \textbf{discontinued further discussion} with the uncivil developer. In five emails, the maintainer \CHANGED{\textbf{escalated the uncivil communication} by fighting for words or accusing the developer's assumptions. In four emails, the maintainer \textbf{discussed in a civil way} with the developer, by providing a technical explanation or trying to understand the problem further.} In three emails, the maintainer \textbf{reinforced their standpoint}, stating that the developer should respect the convention or the workflow. \CHANGED{In three emails, the maintainer \textbf{provided technical explanation} to the developer about the topic in a civil way.} In two emails the maintainer accepted what the developer suggested and \textbf{made a compromise} with the developer. In \CHANGED{other} two emails, \CHANGED{the developer \textbf{accepted the maintainer's criticism} and addressed the changes suggested by the maintainer in the source code. Finally, in one email, the maintainer \textbf{tried to stop the incivility} after a long fight.}

\subsubsection{Consequences of incivility in maintainers' emails}
We found \CHANGED{eight} different categories of consequences when maintainers are uncivil. Very frequently (in \CHANGED{24} cases), the developer \textbf{discontinued further discussion} on the topic by not replying to the email or abandoning the patch. Those cases are ``silent rejects'' of the submitted patches. In \CHANGED{19} cases, the developer simply \textbf{accepted the maintainers’ criticism} and performed the requested changes based on the maintainers', however uncivil, feedback. In \CHANGED{18} cases, the developer \textbf{discussed in a civil way} with the maintainer, by providing technical explanations, discussing alternative solutions, asking for clarifications, or trying to reach a consensus with the maintainer. \CHANGED{In ten cases}, some developers \textbf{escalated the uncivil communication}, attacking the maintainer back in an uncivil way. \CHANGED{In nine emails, the developer \textbf{provided technical explanation} about the change or the problem in a civil way. Rarely (in three cases)}, either the developer or a third party (e.g. another maintainer) \textbf{called out the uncivil behavior}, asking for more constructive feedback or change of maintainer. \CHANGED{Even rarer (in two cases), the developer \textbf{reinforced their standpoint}, and in only one case, the developer \textbf{made a compromise} with the maintainer.}

\subsubsection{Cause and consequence relationship}

\begin{figure}[t]
\centering
\includegraphics[clip, width=\linewidth]{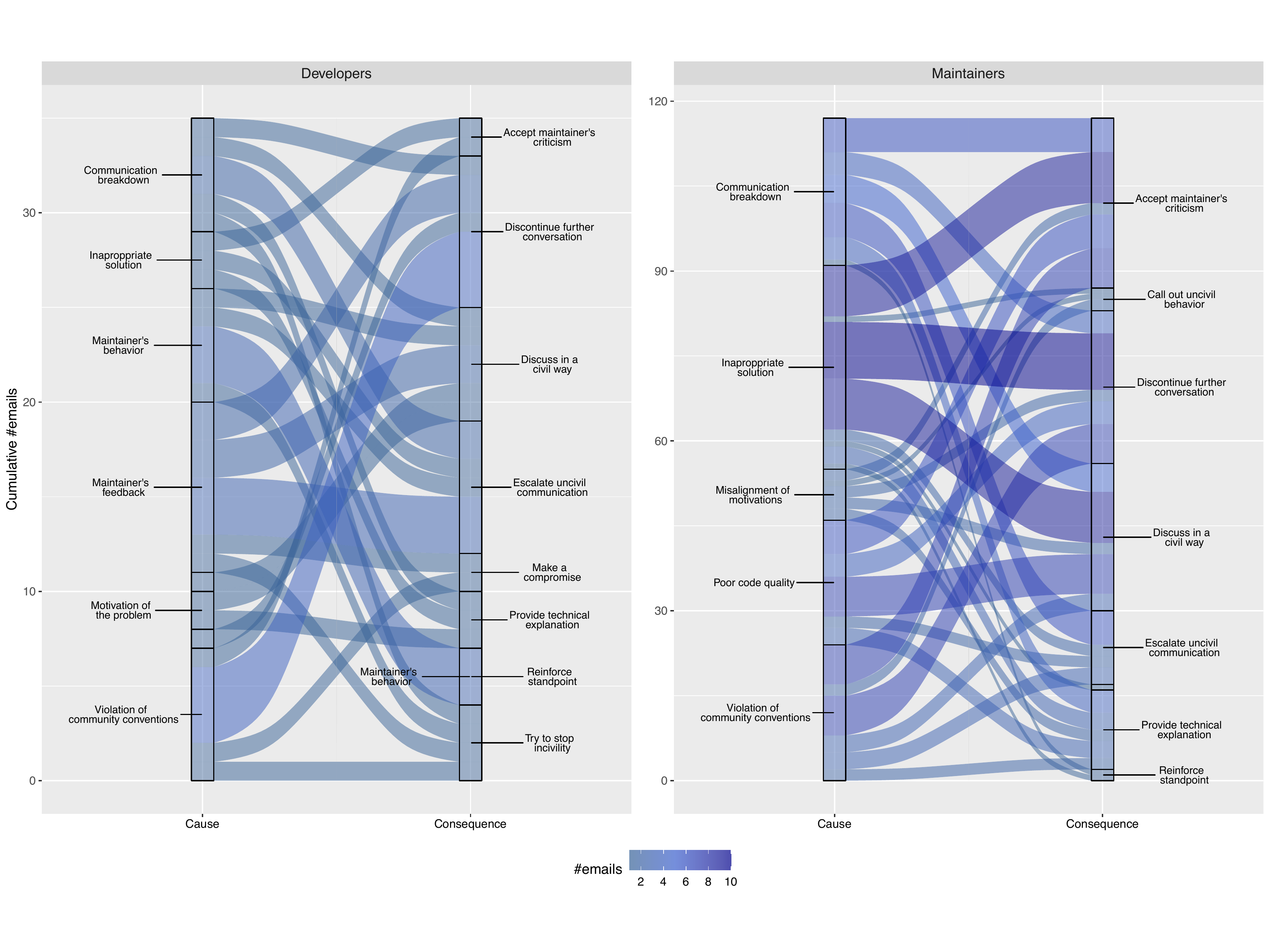}
\caption{\CHANGED{Relationship between causes and consequences of uncivil emails sent by developers (left) and maintainers (right) \CHANGEDFINALVERSION{when discussing rejected patches.}} 
}
\label{fig:cause-impact-dev-maintainers}
\end{figure}

We also assessed the relationship between the identified causes of RQ4 and the consequences of RQ5. %The left-hand side of the diagram presents the causes, and the consequences are shown in the right-hand side. As a result, 
Figure~\ref{fig:cause-impact-dev-maintainers} \CHANGED{(left)} summarizes this relationship in uncivil emails sent by developers. We observe that when developers sent uncivil emails due to \textit{rejection}, \textit{violation of community conventions}, or \CHANGED{\textit{maintainer's feedback}}, the maintainer most likely discontinued the discussion.  %, and in some cases provided a technical explanation, made a compromise with the developer, or tried to stop the incivility. 
When the cause of incivility was the \CHANGED{\textit{maintainer's feedback}, \textit{communication breakdown}, \textit{maintainer's behavior}, or \textit{innapropriate solution}, the maintainer most likely escalated the uncivil communication. Finally, when developers were uncivil due to the \textit{maintainers' behavior}, the maintainer most likely reinforced their standpoint.}%In some cases, the maintainer escalated the uncivil communication, tried to stop the incivility, or discontinued further discussion. Finally, when the cause was a \textit{communication breakdown}, the maintainer often discontinued the conversation. In few cases, the maintainer tried either to stop the incivility or to escalate the uncivil communication.
%Finally, when the cause was the \textit{maintainer behavior}, the maintainers often reinforced their standpoint.

Conversely, when maintainers sent uncivil emails (Figure~\ref{fig:cause-impact-dev-maintainers} \CHANGED{(right)}), in most of the cases, regardless of the cause of the maintainer's uncivil behavior, the developer often accepted the maintainers' criticism, discontinued further conversation, or discussed the problem in a civil way. The developers only escalated the uncivil communication when the cause was \CHANGED{mostly} a \textit{communication breakdown}. Finally, the developers usually called out the uncivil behavior because the behavior was caused by a \textit{violation of community conventions} or \CHANGED{\textit{poor code quality}}. %In a few cases, the developer escalated the uncivil communication or called out an uncivil behavior. When the cause for maintainers was the \textit{inappropriate solution proposed by developer}, the developer often discontinued the conversation. In some cases, the developer accepted the maintainers' criticism or discussed the problem in a civil way. Finally, When the cause was \textit{communication breakdown}, developers discontinued the conversation, discussed in a civil way, or accepted the maintainers' criticism.\\

% \begin{figure}[t]
% \centering
% \includegraphics[clip, width=\linewidth]{images/cause_consequence_dev_maintainer.pdf}
% \caption{\CHANGED{Relationship between causes and consequences of uncivil emails sent by developers (left) and maintainers (right) \CHANGEDFINALVERSION{when discussing rejected patches.}} 
% }
% \label{fig:cause-impact-dev-maintainers}
% \end{figure}

% \begin{figure}[t]
% \centering
% \includegraphics[clip, width=0.725\linewidth]{images/cause_impact_dev.png}
% \caption{\CHANGED{Relationship between causes (left) and consequences (right) of uncivil emails sent by developers}}
% \label{fig:cause-impact-dev}
% \end{figure}

% \begin{figure}[t]
% \centering
% \includegraphics[clip, width=0.725\linewidth]{images/cause_impact_reviewer.png}
% \caption{\CHANGED{Relationship between causes (left) and consequences (right) of uncivil emails sent by maintainers}}
% \label{fig:cause-impact-maintainer}
% \end{figure}

\section{Discussion}
In this section, we discuss the main findings of our analysis, propose practical approaches and research directions for addressing incivility in the software development context, and scrutinize techniques for incivility detection.

\subsection{Discussion on the main findings}

\textbf{Our results show that incivility is common in code review discussions of rejected patches.} We found that \CHANGED{66.66\%} of the non-technical emails, \CHANGED{which corresponds to 7.25\% of all analyzed emails}, are uncivil. 
%This result is similar to what was found by Coe et al.~\cite{coe2014online}, in which 55.5\% of newspaper discussions were uncivil \bram{but that was on all newspaper discussions, while here our percentage is relative to non-technical emails only}.
We found that the most common types of tone-bearing discussion feature (TBDF) in uncivil comments were \textit{bitter frustration}, \textit{name calling}, and \CHANGED{\textit{impatience}}. Although our identification of uncivil TBDF was inspired by the work of Coe et al.~\cite{coe2014online}, we have found additional uncivil features in our context, i.e., \textit{bitter frustration}, \CHANGED{\textit{impatience}}, \textit{irony}, \CHANGED{\textit{mocking}, and \textit{threat}}. Also different from Coe et al., we did not encounter \textit{aspersion}, \textit{lying accusation}, and \textit{pejorative for speech} in our data. We speculate that these differences from Coe et al.'s work have reflected the special nature of discussion in code reviews. Code review discussions are often lengthy and extensive, resulting in frustration if agreement cannot be achieved. Moreover, such discussions are usually warrant-based (i.e., rely on laying out rationales and beliefs), rather than evidence-based (i.e., rely on the accuracy of factual supports)~\cite{Wang2020}, resulting in more confrontational discussion features and less accusation for lying.

\textbf{Against our expectations, we did not find evidence that incivility is related to arguments.} Although we found that code review email threads with an argument tend to have longer discussions, we did not find evidence that discussions containing arguments included more uncivil communication. Hence, given the nature of code review that tends to have long discussions in which participants tend to disagree~\cite{angouri2012theorising}, our results indicate that people can still disagree in a civil way. Moreover, we have found that there was no argument in the threads that contained about half of the uncivil emails.
%in about half of the uncivil email threads, there was no argument. 
In many cases, the other party stopped the communication facing incivility. %\bram{how many people does this correspond to, i.e., would this save a lot of people?}. 
This finding echoes results from previous work~\cite{Huang2016} that found conflicts to cause members to leave the project. Consequently, our results support the notion that open source communities might be able to retain more contributors by fostering civil arguments. Concretely, by avoiding the expression of the uncivil TBDFs identified in this study (e.g., bitter frustration, name calling, and impatience), code review discussion participants may make more constructive and efficient arguments.

% Concretely, based on the uncivil TBDFs we identified, if code review discussion participants cease the expression of \textit{bitter frustration}, \textit{name calling}, and \CHANGED{\textit{impatience}}, reviews and arguments could be more constructive and efficient.

%\CHANGED{and developers would be more productive, since frustration may lead to poor outcomes and negative learning performance~\cite{Denae2015}.}

Our results also show that only four contributors have sent only uncivil emails and 54 contributors have sent not only uncivil emails but also civil and technical emails. \textbf{Moreover, we could not find evidence that incivility is correlated with the authors of uncivil emails.
%a few contributors sent emails containing uncivil TBDFs.
} Concerning the topic of the discussion, even though \textit{workflow} is the most common discussion topic that contained uncivil emails sent by developers, and \CHANGED{\textit{system components} the corresponding topic in uncivil emails sent by maintainers}, \textbf{we could not find evidence that there is a correlation between topics and incivility}. Hence, \textbf{uncivil comments can potentially be made by any people when discussing any topic}. This result suggests that contributors should be mindful when writing or replying to review emails, since previous work has found that lack of respect can create negative perceptions for contributors as well as hinder collaboration~\cite{Bosu2017}.

Since we could not find evidence that incivility is related to common assumptions, we then assessed the causes and consequences that are visible in public code review discussions. \textbf{We found that developers were uncivil mostly because of  \textit{the maintainers' feedback}, \textit{violation of community conventions}, and \textit{communication problems}.} The consequences of these uncivil comments on maintainers are diverse. While maintainers often simply discontinued the discussion, they most frequently followed up with a civil discussion on the technical level, if a response was provided. \CHANGED{These results are similar to the ones found in our motivational case study (Section~\ref{sec:case-study}), in which participants mentioned that civility is related to \textit{constructive feedback}}. \CHANGED{However,} %although the effort \CHANGED{of having a civil discussion} does not decrease the amount of incivility in the discussion, 
previous work~\cite{Huang2016} has found that, in general, pure technical explanations have no effect on retaining contributors, since those explanations are often superficial and demonstrate a misunderstanding of the contributors' work. In other cases, maintainers \textit{reinforced their standpoint}, \textit{escalated the uncivil communication}, or \textit{tried to stop the incivility}. \CHANGED{This was also mentioned in our case study, in which participants said that contributors might escalate the problem if nobody recedes.} Even when maintainers reinforced their standpoint or tried to stop the incivility, the tone in which the feedback is delivered can cause unnecessarily interpersonal conflicts~\cite{egelman2020predicting}. 

\textbf{When maintainers are uncivil, mostly due to a developer's \textit{violation of community conventions}, \textit{inappropriate solution proposed by developer}, or \textit{communication breakdowns}, developers often \textit{accepted the maintainers' criticism} and \textit{discussed the problem in a civil way}.} One reason for that might be the power imbalance between maintainers and developers that have resulted in developers' polite resistance to blame maintainers for uncivil communication or to fight back. Although rare, a few developers \textit{escalated the uncivil communication} and \textit{called out uncivil maintainers}. We also observed that very frequently developers \textit{discontinued further conversation} by not replying to the uncivil maintainer or simply abandoning the patch. To avoid this to happen, maintainers should keep in mind that open source contributors may need to be intrinsically motivated such as by feeling competent and being understood~\cite{Huang2016, alexander2002working}.

\CHANGED{\subsection{Proactive and reactive approaches to address risk factors before and after incivility happens}}
\label{sec:proactive-implications}

Based on our results, we propose some practical implications and suggestions for open source communities \CHANGED{and researchers}. \CHANGED{We split the implications into \textit{proactive approaches}, i.e., what can be done to address the causes of incivility and to identify potential risks before uncivil communication happens, and \textit{reactive approaches}, i.e., what can be done to identify and address incivility after it happens.}

\paragraph{\CHANGED{\textbf{Proactive approaches.}}} 
\CHANGED{Our study has identified several frequent causes of incivility. We argue that, if evident, open source software (OSS) communities should first focus on addressing these causes in order to remove factors that may result in uncivil communication in the first place. For each of these causes, we propose in Table~\ref{tab:proactive-implications} some practical approaches for both OSS communities and researchers to address.}

\begin{table}[ht]
\centering
\caption{Proactive approaches for OSS communities and researchers.}
\label{tab:proactive-implications}
\small
\scalebox{0.57}{
\begin{tabular}{llll}
\toprule
\textbf{Role}  & \textbf{\begin{tabular}[c|]{@{}l@{}}Most frequent \\ causes of incivility\end{tabular}}  & \textbf{Practical approaches for OSS communities} & \textbf{Practical approaches for researchers}\\ \midrule

\begin{tabular}[c|]{@{}l@{}}Developers \&\\ Maintainers\end{tabular}  & \begin{tabular}[c|]{@{}l@{}}Violation of\\ community\\ conventions\end{tabular} &
\begin{tabular}[c|]{@{}l@{}} \tabitem Include a training for newcomers and developers to ensure \\ that everyone is aware about the community conventions, \\especially if the conventions change.\\ \tabitem Maintainers should always include why the patch was rejected.\\ Violation of community conventions should not be a reason for\\ silent rejection. \\ \tabitem Gamify the review process so that developers that \\ follow the community conventions gain more reputation and status.\end{tabular}  & \begin{tabular}[c]{@{}l@{}} \tabitem Develop tools that help developers in the review process.\\  For example, if a cover letter is mandatory (see example \\ in Section~\ref{sec:causes}) and the developer forgot to add it, then the \\ tool would warn the developer that something is missing\\ before the message is sent.\end{tabular} \\ \midrule

\begin{tabular}[c|]{@{}l@{}}Developers \&\\ Maintainers\end{tabular} & \begin{tabular}[c|]{@{}l@{}}Communication \\ issues\end{tabular}  & \begin{tabular}[c|]{@{}l@{}} \tabitem Develop a code of conduct~\cite{tourani2017code} focused on the code review process \\ by providing guidelines on how to communicate constructive feedback\\ (maintainer's side) and how to interpret the feedback (developer's side).\end{tabular}   & \begin{tabular}[c|]{@{}l@{}} \tabitem To avoid poorly articulated explanations and arguments,\\ researchers could develop tools that help linking technical \\ explanations to the relevant code snippets in order to make \\ the discussion more evidence-based and the arguments \\ more effective.\end{tabular}            \\ \midrule

Developers & \begin{tabular}[c]{@{}l@{}}Maintainer's \\ feedback\end{tabular}    & \begin{tabular}[c|]{@{}l@{}} \tabitem Include a training for maintainers on how to give constructive feedback~\cite{alami2019does}.\\ \tabitem Include a training for developers on how to handle rejections so that they \\ are aware that rejection is not a failure~\cite{alami2019does}.\\ \tabitem Make coaching or mentoring sessions available for maintainers~\cite{alami2019does}.\\ 
%\tabitem Gamify the review process so that maintainers that give constructive \\ feedback in a civil way gain more reputation and status.
\end{tabular} 
& \begin{tabular}[c|]{@{}l@{}} \tabitem Develop tools for supporting maintainers to give constructive \\ feedback.  \\  \tabitem Develop strategies to gamify the review process so that \\ maintainers that give constructive feedback in a civil way gain \\ more reputation and status. \end{tabular} \\ \midrule

Maintainers &\begin{tabular}[c|]{@{}l@{}} Inappropriate \\solution\end{tabular}     & \begin{tabular}[c]{@{}l@{}} \tabitem Developers should always include a technical rationale of their solution,\\ including the negative side effects of the solution (if there are any), the \\ motivation of the proposed patch, and the limitations.\\ \tabitem Provide awareness to developers in the sense that even if the solution is not\\ appropriate, the code review practice enables to promote knowledge \\ sharing and learning opportunities~\cite{alami2019does}, and there is no need to discontinue \\ further conversation.\end{tabular} &
\begin{tabular}[c|]{@{}l@{}} \tabitem Researchers could survey or interview OSS developers and \\ assess the extent to which developers accept the maintainers' \\ criticism due to power imbalance, and what are the \\ consequences for OSS communities of just accepting the criticism \\ without further interaction.\end{tabular} \\ \midrule

Maintainers & Poor code quality & \begin{tabular}[c|]{@{}l@{}} \tabitem Include a training for newcomers and developers to ensure that everyone \\ is aware of the community's expectations in terms of code quality.\\ \tabitem Adopt existing code analysis tools, integrating them into the \\ developers' workflow (e.g., continuous integration).  \end{tabular}  & \begin{tabular}[c|]{@{}l@{}} \tabitem Develop tools to support developers by checking for code\\ quality (such as readability and performance) before the patch \\ ends in the mailing list.\end{tabular}   \\ \bottomrule                 
\end{tabular}}
\end{table}

\CHANGED{In addition to addressing the causes, other approaches may help contributors to avoid posting uncivil comments on the mailing list. For example, contributors could use tools to check if their emails are uncivil before they are sent to the mailing list. A more fine-grained tool that lets contributors know what kind of incivility is present in their email would also help them to change their text for a more civil discussion. In the future, these types of tools could be integrated into the code review process, yet they will rely on automated or semi-automated techniques for the detection of incivility in potential comments. We discuss these techniques in Section~\ref{sec:incivility-detection}}.

\paragraph{\CHANGED{\textbf{Reactive approaches.}}} 

\CHANGED{Although \textit{proactive approaches} help OSS communities to prevent incivility, they do not take into consideration the cases when incivility has already happened in the open. For that, \textit{reactive approaches} need to be considered. That is, when incivility happens, community leaders need to do damage control, and community members need to be informed and properly respond to the incivility. We argue that if OSS communities implement both approaches in practice, incivility can be considerably reduced.

To identify and address incivility after it happens, OSS communities could use a bot that is constantly checking if the emails sent to the mailing list are civil or uncivil. A crowd-based technique can also be investigated to allow community members to collaboratively identify uncivil conversations in code reviews and augment the automated tools. If incivility keeps happening in an email thread, community leaders can be warned to assess the situation and take the appropriate measures, such as applying the code of conduct. In case the OSS community already uses bots to identify ``heated conversations'', such as the Stack Overflow bot\footnote{\url{https://github.com/SOBotics/HeatDetector}}, the community could then incorporate the TBDFs found in this study into their tool. This will allow OSS communities to identify more cases and types of incivility, and to % . That is, instead of only returning a score of how ``heated'' the discussion is (as the current bots are doing), one could do better and make these bots identify the type of incivility, such that the community would provide awareness of what type of unnecessary communication is happening, and it would allow them to
target and mitigate specific types of incivility more effectively. Similar to some of the proactive approaches, these reactive approaches also can benefit from techniques for the detection of uncivil comments, which we discuss next.}

\subsection{Incivility detection}
\label{sec:incivility-detection}

%[Mention again that some of the approaches above rely automated incivility detection]

\CHANGED{As we have previously discussed, some of the proactive and reactive approaches to address incivility before and after it happens rely on automated incivility detection. For that, three approaches could be considered. First, sentiment analysis tools are commonly used to evaluate whether a conversation is \textit{positive}, \textit{negative}, or \textit{neutral}. Although current sentiment analysis tools do not identify incivility, they could provide hints of whether a conversation is negative or not. Second, toxicity and offensive language tools could be used to identify expressions whose intention is to harm other people. Finally, incivility could be automatically identified through the TBDFs found in this study. We discuss below how the aforementioned approaches could be addressed.}

\paragraph{\CHANGED{\textbf{Sentiment analysis tools.}}} \CHANGED{In our motivational case study (see Section~\ref{sec:case-study}), we compared the results of sentiment analysis tools with the perception of Linux developers. Although we had a very small sample (three emails), we observed that there was a lack of agreement between sentiment analysis tools (Senti4SD and IBM Watson), and between tools and humans. Additionally, based on the feedback received in our case study and on our own experience conducting the qualitative analysis on our dataset, we speculated that current sentiment analysis tools might not be able to identify incivility. To confirm this speculation, we extended our case study by analyzing the sentiment of all sentences in our dataset coded with a TBDF, adding up to 337 distinct sentences. Our goal is to assess if the existing sentiment analysis tools are able to detect incivility.}

\CHANGED{\paragraph{Methods.}
To achieve our goal, we run three software engineering (SE) specific tools to detect sentiment, namely Senti4SD~\cite{calefato2018sentiment}, SentiStrength-SE~\cite{islam2018sentistrength}, and SentiCR~\cite{ahmed2017senticr}. We only considered SE-specific tools because previous research~\cite{novielli2015challenges, jongeling2017negative, lin2018sentiment} has found that general-purpose sentiment analysis tools need to be fine-tuned to accommodate the technical-heavy discussions in the software development context. Further, we decided to use pre-trained models because this is an exploratory study and we might not have a dataset that is big enough to train a classifier. Moreover, we chose to compare the results of the three aforementioned tools for the following reasons.

% Furthermore, previous research~\cite{novielli2020can} has found that cross-platform sentiment analysis tools, that is, tools trained and tested in different datasets, might not perform well. In the best case scenario, it is recommended to retrain SE-specific tools, but for that a minimum training set of about 1,000 documents is required~\cite{novielli2020can}. Since, to the best of our knowledge, we are the first ones to create an incivility dataset in the SE domain, and we only have 337 sentences coded with a TBDF, retraining a tool is not an option for us. 

First, we chose Senti4SD~\cite{calefato2018sentiment}, a supervised tool trained and validated on 4,000 questions, answers, and comments from StackOverflow, because it is the tool that has achieved the best performance when compared to other tools~\cite{calefato2018sentiment, novielli2020can, novielli2020assessment} and it reduces misclassifications of neutral and positive posts as emotionally negative~\cite{novielli2020can}. Second, we chose SentiCR~\cite{ahmed2017senticr} because it is the only SE-specific tool trained on code review comments from Gerrit. Additionally, SentiCR performs the SMOTE~\cite{chawla2002smote} technique to handle class imbalance in the training set (similar to our case). Finally, we chose SentiStrength-SE~\cite{islam2018sentistrength} because it implements a lexicon-based approach. Although SentiStrength-SE is trained on issue comments from Jira, the tool is unsupervised; unsupervised tools are found to perform better than supervised tools when retraining is not possible~\cite{novielli2020can}. Since different tools return different sentiment polarity labels as an output, we converted the outputs into \textit{positive}, \textit{negative}, and \textit{neutral} based on the mapping suggested by the literature~\cite{novielli2020can}, except for the SentiCR tool, which that only returns the \textit{negative} and \textit{non-negative} polarities. We used the sentences manually labeled in our dataset as gold standard. Then, we converted the TBDFs into sentiment polarities, as described in Section~\ref{sec:rq1-results}. Since current sentiment analysis tools do not detect incivility, we consider the uncivil TBDFs as having a \textit{negative} sentiment.
%, while \bram{what about positive/neutral? It's described in the previous sentence}. 

To assess the performance of the tools, we computed the typical classification metrics: precision, recall, and f-score and calculated their micro- and macro-averages. The \textit{precision}~\CHANGEDFINALVERSION{\cite{buckland1994relationship}} for a given sentiment polarity (e.g., positive) was calculated as the ratio of sentences for which a given SE-specific sentiment analysis tool correctly identified the presence of that polarity% to the number of sentences labeled with the sentiment polarities by the tools
. The \textit{recall}~\cite{buckland1994relationship} for a given polarity is the ratio of all sentences with that polarity that a given SE-specific sentiment analysis tool was able to find. \textit{F-score }is the harmonic mean of the precision and the recall. For completeness, we also report the overall performance using micro-averaging and macro-averaging as aggregated metrics~\cite{sebastiani2002machine}. Micro-averaging is influenced by the performance of the majority polarity class, and macro-averaging is mostly used when the dataset is unbalanced, since it accounts for the classifier's ability to identify classes with few datapoints. %\bram{OK, but how does one calculate micro/macro-averaging?} 
Since our data is unbalanced, we will mostly rely on the macro-averaging values when globally analyzing our results, and on the precision and recall values when analyzing the results by sentiment polarity.}

\begin{figure}[ht]
\centering
\includegraphics[clip, width=0.9\linewidth]{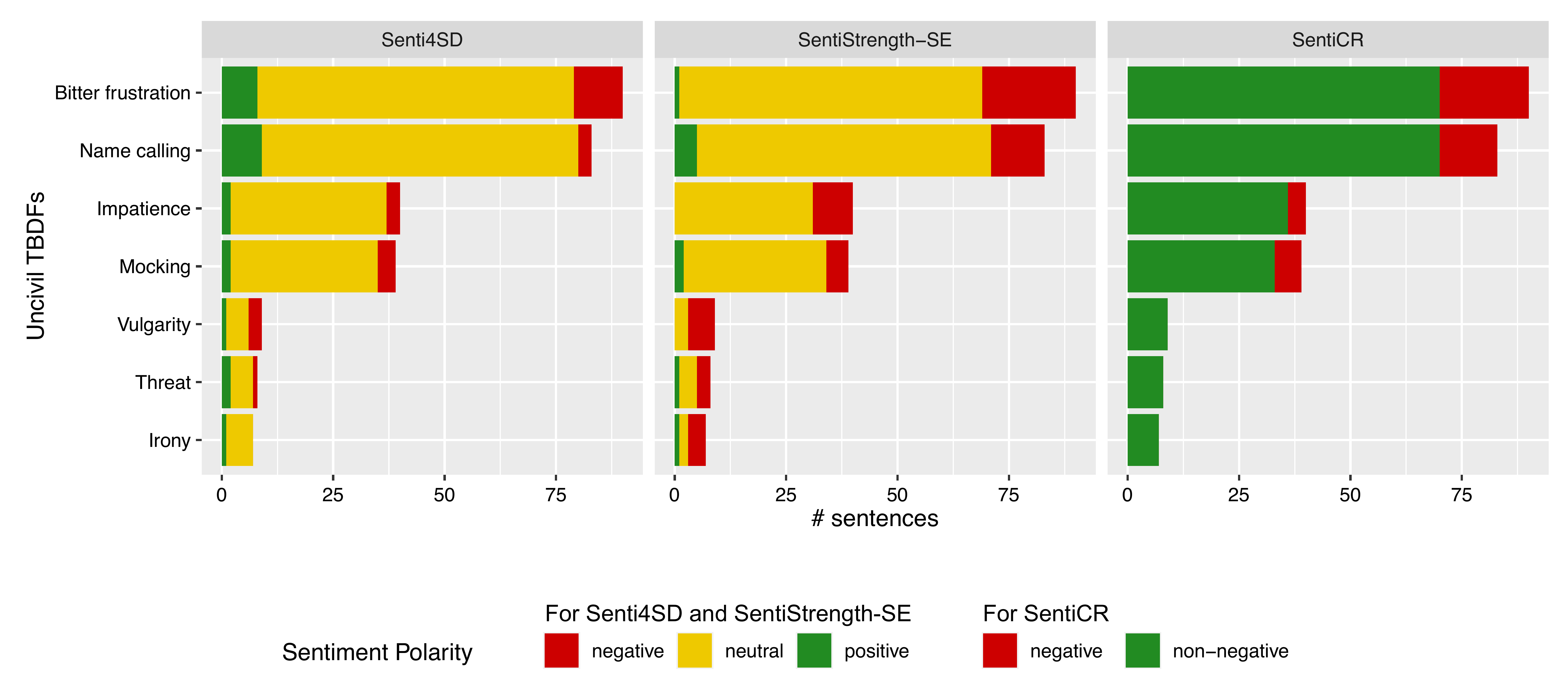}
\caption{\CHANGED{Sentiment polarity of uncivil TBDFs.}}
\label{fig:uncivil-sentiment}
\end{figure}

\paragraph{Results.}
\CHANGED{\textbf{The three SE-specific sentiment analysis tools tend to have high precision for the positive and negative classes, and high recall for the neutral class. However, the overall performance (F1, micro and macro-averaging) is very low for all analyzed tools} (see Table~\ref{tab:sentiment_tools_performance}). Furthermore, we observe in Figure~\ref{fig:uncivil-sentiment} that most tools classified sentences coded with an uncivil TBDF as \textit{neutral} or \textit{non-negative}, which explains the low recall for the negative class. According to Novielli et al.~\cite{novielli2020can}, it is expected to find a drop in precision for the neutral class, and recall for the negative and positive classes when analyzing the results in a cross-platform setting, i.e., training and test sets are different. Previous work~\cite{novielli2020can} has also found that this might be due to the fact that positive and negative lexicons are platform-dependent.

Based on these results, we conclude that current SE-specific sentiment analysis tools do not perform well when detecting incivility. In particular, while a sentiment analysis tool might be relatively convincing when identifying an uncivil message (negative sentiment; precision of 73\% to 77\%), it would miss up to 91\% of those cases (Senti4SD). In fact, incivility has many dimensions that are not captured by sentiment analysis tools, such as the context of the conversation (in our case the emails prior to the uncivil email in a thread), the familiarity among people, and the granularity of analysis. Furthermore, some TBDFs are not sentiment-related, such as \textit{irony}, \textit{mocking}, and \textit{threat}, and it might be hard to capture them with sentiment models only. Hence, with current technologies, \textbf{incivility cannot be captured reliably only by analyzing the sentiment of a text.}}
  
% \bram{I don't understand the link with previous analysis, sounds like we just discussed that with precision/recall? better to integrate this paragraph in discussion of precision/recall as a means to understand why recall is so low for negative}
  
% \CHANGED{Finally, we analyze if sentiment analysis tools are able to classify sentences coded with an uncivil TBDF as \textit{negative}. As presented in Figure~\ref{fig:uncivil-sentiment}, \bram{following sounds trivial, since a low recall means exactly that, not sure if we need that figure} 
%  As a conclusion, we found that \textbf{sentiment analysis tools are not able to detect uncivil TBDFs}. \bram{integrate in previous paragraph:} 

\begin{table}[ht]
\centering
\small
\aboverulesep=0ex
\belowrulesep=0ex
\caption{\CHANGED{Performance of SE-specific sentiment analysis tools. For each tool, we highlight the best values for each metric.}}
\label{tab:sentiment_tools_performance}
\begin{tabular}{l|ccc|ccc|ccc}
\toprule
\multicolumn{1}{c|}{\multirow{2}{*}{\begin{tabular}[c|]{@{}c@{}}\textbf{Sentiment} \\ \textbf{Polarity}\end{tabular}}} & \multicolumn{3}{c|}{\textbf{Senti4SD}} & \multicolumn{3}{c|}{\textbf{SentiStrength-SE}} & \multicolumn{3}{c}{\textbf{SentiCR}} \\ \cline{2-10}
\multicolumn{1}{c|}{}                                                                              & \textbf{Precision}  & \textbf{Recall}  & \textbf{F1}    & \textbf{Precision}     & \textbf{Recall}     & \textbf{F1}      & \textbf{Precision}  & \textbf{Recall}  & \textbf{F1}   \\ \midrule
Negative                                                                                          & \cellcolor[gray]{0.8}0.73       & 0.09    & 0.16  & \cellcolor[gray]{0.8}0.74          & 0.23       & 0.35    & \cellcolor[gray]{0.8}0.77       & 0.16    & 0.26 \\
Neutral                                                                                           & 0.06       & \cellcolor[gray]{0.8}0.70    & 0.11  & 0.05          & \cellcolor[gray]{0.8}0.52       & 0.09    & -          & -       & -    \\
Positive                                                                                          & \cellcolor[gray]{0.8}0.44       & 0.31    & 0.36  & \cellcolor[gray]{0.8}0.65          & 0.29       & 0.40    & -          & -       & -    \\
Non-negative                                                                                      & -          & -       & -     & -             & -          & -       & 0.26       & \cellcolor[gray]{0.8}0.86    & 0.40 \\ \midrule
Micro-averaging                                                                                   & 0.17       & 0.17    & 0.17  & 0.26          & 0.26       & 0.26    & 0.34       & 0.34    & 0.34 \\
Macro-averaging                                                                                   & \cellcolor[gray]{0.8}0.41       & 0.37    & 0.21  & \cellcolor[gray]{0.8}0.48          & 0.34       & 0.28    & \cellcolor[gray]{0.8}0.52       & 0.51    & 0.33\\
\bottomrule
\end{tabular} 
\end{table}

% \CHANGED{\paragraph{Practical implications.} Based on the aforementioned results, we provide some practical implications for researchers.}

% \begin{itemize}
%     \item \CHANGED{Researchers could use our public available\footnote{\url{TODO}}  \hl{TODO} dataset and codebook as a start point to extend our study and analyze more code review emails from the Linux Kernel. With more datapoints, it will be possible to retrain SE-specific tools and assess the performance in the within-platform setting, i.e., training and testing in the same dataset. Only after that it will be possible to know if sentiment analysis tools work for identifying incivility or not.}
    
%     \item \CHANGED{Since the positive and negative lexicons are platform-dependent, researchers could use our dataset to build a lexicon specific to the Linux community.}
% \end{itemize}

\paragraph{\CHANGED{\textbf{Detection of toxicity and offensive language.}}} \CHANGED{Although there are existing general-purpose tools that analyze toxicity in online communication, such as PerspectiveAPI\footnote{\url{http://perspectiveapi.com}} and Tensorflow toxicity model\footnote{\url{https://github.com/tensorflow/tfjs-models/tree/master/toxicity}}, these tools do not capture the broad spectrum of incivility. In fact, previous work defines toxicity in online communities as \textit{``(explicit) rudeness, disrespect or unreasonableness of a comment that is likely to make one leave the discussion''}~\cite{xia2020toxiciy}. Although there is an overlap between toxicity and incivility, toxicity only covers one dimension of incivility, i.e., language that harms other people. Incivility is more general and focuses on issues that can hurt a constructive and technical conversation. According to Sadeque et al.~\cite{sadeque2019incivility}, a fine-grained incivility detection is more challenging than toxicity detection, and the differences between these tasks (incivility and toxicity detection) make it hard to use the same data or the same strategies for both tasks. Furthermore, Hosseini et al.~\cite{Hosseini2017} have found that PerspectiveAPI often identifies false positives (i.e., assign high toxicity scores to sentences that are not toxic), and that the tool classifies a sentence and an adversarial sentence (modified sentences that contain the same highly abusive content as the original one) with completely different toxicity scores. In the SE context, Raman et al.~\cite{ramanstress} have built a combination of general pre-trained sentiment analysis tools and toxicity classifiers, such as PerspectiveAPI. Although their best classifier had a precision of 0.91, the recall was very low (0.42), showing that the tool was not able to identify the majority of unhealthy interactions. On top of that, the classifier shows very low precision (0.50) when tested on random issues. Based on that, we claim that a more fine-grained classification using software development data and the TBDFs found in this study is needed to produce accurate results when detecting incivility.}

\paragraph{\CHANGED{\textbf{Identifying incivility through TBDFs.}}} \CHANGED{Given the fact that the existing sentiment analysis, toxicity, and offensive language tools cannot readily identify uncivil comments, dedicated tools and techniques should be built. The TBDFs proposed in our study can be used as a framework to support these techniques. For example, heuristics could be developed to identify each TBDF, a civility-specific lexicon for each TBDF and for each community could be built to improve the performance of classifiers, and our dataset\footnote{\replicationPackage} could be extended based on the proposed framework with the goal of training machine learning models to detect incivility.
% \hl{[explain some technical directions for classifiers... ontology/heuristic-based approaches (develop heuristics for each TBDF)? build a civility-specific lexicon for each TBDF? use this TBDF framework to extend a larger dataset for training NNs? etc.]} 
Additionally, some existing tools could be extended based on our TBDF framework and dataset. For example, Gachechiladze et al.~\cite{gachechiladze2017anger} have proposed a tool to detect anger and its direction in Apache issue reports; their approach could be extended to identify the tone (e.g. impatience, irony, etc) behind the anger with the goal to explain the reasons behind it.}

\section{Threats to validity}

\CHANGED{In this section, we discuss the major threats to the validity~\cite{wohlin2012experimentation} of our study, in the following categories.}

\CHANGED{\paragraph{\textbf{Construct validity.}}
% Relation between theory and observation

The TBDFs identified in this study might not capture (in)civility in practice. To minimize this threat, we started our analysis with a civility framework~\cite{coe2014online} and we built our work on top of that. Furthermore, the categorization of TBDFs was made with two other authors to avoid biases and misclassifications.
}

\CHANGED{\paragraph{\textbf{Internal validity.}}
% Factors that can affect the results.

 Our qualitative coding could lead to inconsistencies due to its subjectiveness. To minimize this threat, our codebook was iteratively improved based on discussions with two other authors. Additionally, the second author analyzed all emails in which the first author has identified a TBDF, and as a result, we found \CHANGEDFINALVERSION{on average} a substantial agreement between the two raters. \CHANGEDFINALVERSION{The Kappa values varied among the TBDF codes, probably due to the different difficulties inherent to each TBDF; however, we have achieved at least a moderate agreement between the two raters on each individual code.} The main threat of our study concerns the technical emails identified by the first author, which were not verified by a second rater. The technical emails were identified with a straightforward list of criteria (see Section~\ref{sec:qualitative-analysis}), with a low risk of misclassification.

  For the analysis of individual contributors, we assess the number of contributors that have sent (un)civil emails as well as the contributors' roles. To do that, we grouped contributors' aliases either by the same name or the same email. To mitigate the risk of having wrong identities clustered together, we manually checked all clusters to assess their correctness. Finally, the actual size of the population of rejected patches is unknown, since the heuristics used in this paper might have false positives and false negatives. To mitigate this risk, we have assessed the performance of the heuristics in another dataset, since there is no available ground truth for the LKML.
}

\CHANGED{\paragraph{\textbf{Conclusion validity.}}
% Concerns the relation between the treatment and the outcome

Conclusion validity concerns the statistical analysis of the results, in which commonly used statistical techniques are applied to validate the researchers' assumptions~\cite{wohlin2012experimentation}. A common threat to this type of validity includes the low number of samples, which reduces the ability to reveal patterns in the data~\cite{wohlin2012experimentation}. Hence, in the quantitative part of our analysis, we aimed at achieving sufficient analysis reliability. We applied statistical tests to assess the correlations of incivility with email and thread attributes, and we made our conclusions based on the statistical power encountered.}

\CHANGED{\paragraph{\textbf{External validity.}}
% Concerns the generalization of results

  The analysis of an open source project represents a threat to the study validity since open source projects and proprietary projects may have different types of incivility. We focus on open source because it is difficult to have access to code review discussions of proprietary systems. The large amount of publicly available data in open source contexts also allows us to examine the phenomenon of (in)civility on a large scale. Additionally, we only analyze one open source community, the Linux kernel. Linux is a popular and large open source project, and the Linux community is very diverse in terms of expertise, gender, ethnicity, and companies contributing to it. Therefore, it is important to have a healthy community to attract and retain contributors~\cite{steinmacher2014attracting}. However, we do not have evidence to support that the results found in this study are generalizable to other projects.

  Furthermore, despite our efforts in characterizing incivility in code review discussions, we only analyzed review emails of rejected patches in a specific period, and the results found in this study might not be generalizable to accepted patches. However, previous work has shown that more than 66\% of all patches submitted to LKML are rejected~\cite{jiang2013will} and that the Linux community frequently rejects patches using harsh language when reporting the rejection, even though the reasons for rejection are purely technical~\cite{alami2019does}. } %Finally, we were also not able to manually verify all emails collected in the analyzed period. }

\section{Conclusion \& Future Work}

%Our study opens up a new field of research on collaboration and communication in software engineering.

Incivility is an important issue that can potentially affect many open source contributors in various ways. \CHANGED{To the best of our knowledge, this is the first study of an in-depth characterization of incivility in open source code review discussions, providing evidence, descriptions, and explanations of incivility in this dynamic context. By analyzing the code review discussions of the Linux Kernel Mailing List, we encountered TBDFs not previously found in any other study, proposed a definition of incivility based on the uncivil TBDFs, assessed the frequency of incivility, analyzed the correlation with the common assumptions of the cause of incivility (i.e., arguments, contributors, and topics), and assessed the discoursal causes and consequences of developers' and maintainers' uncivil interactions.

As a result, we found that incivility is common in code review discussions of rejected patches of the Linux kernel. We also found that frustration, name calling, and \CHANGED{impatience} are the most frequent features in uncivil emails. Besides that, our results indicate that there are civil alternatives to address arguments, and that uncivil comments can potentially be made by any people when discussing any topic. Finally, we found the main causes and consequences of uncivil communication for both developers and maintainers. 

Previous work have found that interpersonal conflicts and toxicity are rare on Google's code review discussions and on GitHub projects with ``too heated'' conversations~\cite{egelman2020predicting, ramanstress}, but they have negative consequences when they occur. Based on this evidence, we decided to first characterize incivility by analyzing the Linux community, which has been criticized for using harsh language, giving frequent rejections, and negative feedback~\cite{alami2019does}. %As a result, we could better characterize the problem and provide evidence that incivility is a real problem.
Because software development is essentially a communication-intense activity, incivility can arise in any community and development stage. However, the results found in this study may not be generalized to other communities or other software development activities. Hence, we suggest that future research investigate the potential generalizability of our findings in other open source and industrial projects. Additionally, not all causes and consequences of incivility are visible in public code review discussions. An in-depth investigation of the community members' experience and perception of incivility would be helpful to address this problem.

We believe that the findings of this work will pave the way for further studies in software development that aim to analyze incivility and promote civil communication. More specifically, the causes and consequences of incivility found in this study are crucial for devising strategies to handle incivility during code review. Even though many approaches exist to prevent incivility, such as the code of conduct~\cite{tourani2017code}, these approaches do not treat the root of the problem. For example, even though the Linux community has a code of conduct\footnote{\url{https://www.kernel.org/doc/html/latest/process/code-of-conduct.html}}, incivility is still present in their code review discussions. \CHANGEDFINALVERSION{Therefore, it is crucial to investigate other means to tackle the problem of uncivil communication in the context of software development. These efforts can be effectively inspired and informed by this categorization study.}}

%To conclude, this is the first time that the concept of incivility is comprehensively characterized in software development communications.} In future work, researchers could assess if incivility can also be encountered in other open source communities and in accepted patches. Additionally, not all causes and consequences might be visible in public code review discussions. To address this problem, researchers could do surveys and interviews with community members to understand other causes and consequences. Finally, researchers could design tools to promote awareness when incivility occurs.

\section*{Acknowledgements}
The authors would like to thank Kate Stewart (Linux Foundation), Shuah Khan (Linux Foundation), and Dr. Daniel German (University of Victoria) for their valuable insights. The authors also thank the Natural Sciences and Engineering Research Council of Canada for funding this research through the Discovery Grants Program.

\bibliographystyle{ACM-Reference-Format}
\bibliography{bibliography.bib}

% \vspace{0.5cm}
% \textbf{Article 353:}
% \received{October 2020}
% \received[revised]{April 2021}
% \received[revised]{July 2021}
% \received[accepted]{July 2021}

\end{document}